\documentclass[12 pt,draftcls]{IEEEtran} \onecolumn
\ifCLASSINFOpdf
\else
\fi
\usepackage[dvips]{graphicx}
\usepackage[cmex10]{amsmath}
\usepackage{amssymb}
\usepackage{subfigure}
\usepackage{algorithmic}
\usepackage{array}
\usepackage{mdwmath}
\usepackage{eqparbox}
\usepackage[font=footnotesize]{subfig}
\usepackage{graphicx}
\usepackage{algorithmic}
\usepackage{algorithm}
\usepackage{color}

\begin{document}
%
\title{Dynamic Bit Allocation for Object Tracking in Bandwidth Limited Sensor Networks}
%
%
%

\author{Engin~Masazade,~\IEEEmembership{Member,~IEEE,}
        Ruixin~Niu,~\IEEEmembership{Senior Member,~IEEE,}
        and~Pramod~K.~Varshney,~\IEEEmembership{Fellow,~IEEE}
\thanks{E. Masazade and P. K. Varshney are with the Department
of Electrical Engineering and Computer Science, Syracuse University, NY, 13244, USA, e-mail:\{emasazad, varshney\}@syr.edu. R. Niu is with the Department of Electrical and Computer Engineering, Virginia Commonwealth University, Richmond, VA 23284, USA. Email: rniu@vcu.edu. }
\thanks{This work was supported by U.S. Air Force Office of Scientific Research (AFOSR) under Grant FA9550-10-1-0263.}
\thanks{Part of this work was presented at the Fusion'11 conference held at Chicago, IL, July 5-8, 2011.}}

\markboth{Journal of \LaTeX\ Class Files,~Vol.~X, No.~X, September~20XX}%
{Masazade \MakeLowercase{\textit{et al.}}: Bare Demo of IEEEtran.cls for Journals}
%



\maketitle

\begin{abstract}
In this paper, we study the target tracking problem in wireless sensor networks (WSNs) using quantized sensor measurements under limited bandwidth availability. At each time step of tracking, the available bandwidth $R$ needs to be distributed among the $N$ sensors in the WSN for the next time step. The optimal solution for the bandwidth allocation problem can be obtained by using a combinatorial search which may become computationally prohibitive for large $N$ and $R$. Therefore, we develop two new computationally efficient suboptimal bandwidth distribution algorithms which are based on convex relaxation and approximate dynamic programming (A-DP). We compare the mean squared error (MSE) and computational complexity performances of convex relaxation and A-DP with other existing suboptimal bandwidth distribution schemes based on  generalized Breiman, Friedman, Olshen, and Stone (GBFOS) algorithm and greedy search. Simulation results show that, A-DP, convex optimization and GBFOS yield similar MSE performance, which is very close to that based on the optimal exhaustive search approach and they outperform greedy search and nearest neighbor based bandwidth allocation approaches significantly. Computationally, A-DP is more efficient than the bandwidth allocation schemes based on convex relaxation and GBFOS, especially for a large sensor network.
\end{abstract}



%
\IEEEpeerreviewmaketitle

\section{Introduction}
\begin{figure}[hbt]
\centerline{ \begin{tabular}{c}
\includegraphics[width=.6\textwidth,height=!]{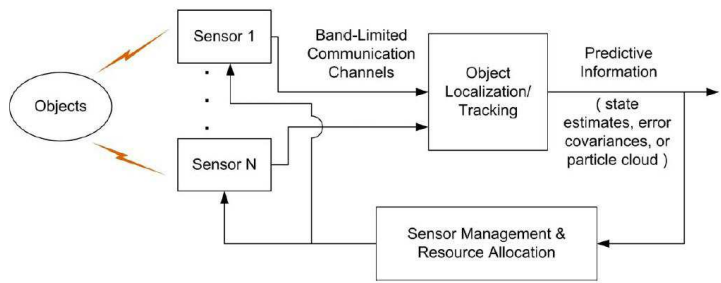} \\
\end{tabular}}
      \caption{System model for sensor and resource management based on feedback from recursive estimator.}
  \label{fig:senman}
\end{figure}

A wireless sensor network (WSN) consists of a large number of spatially distributed sensors which are tiny,
battery-powered devices, and have limited on-board energies. When properly programmed and networked, WSNs perform
different tasks that are useful in a wide range of applications such as battlefield surveillance, environment and
health monitoring, and disaster relief operations. Dense deployment of sensors in the network introduces redundancy in coverage, so selecting a subset of sensors may still provide information with the desired quality. As shown in  Fig. \ref{fig:senman}, the adaptive sensor management policies select a subset of active sensors to meet the application requirements in terms of quality of service while minimizing the use of resources. In this paper, we assume that the task of the WSN is to track a moving target in a given region of interest (ROI). Sensors receive observations from an object of interest and send quantized information to the fusion center over bandwidth limited channels. So the fusion center needs to distribute the available bandwidth among sensors using predictive information based on the target dynamics and the received sensor data. We consider a myopic (one-step ahead) scenario, where at a given time step, the fusion center only decides on the bandwidth distribution of the next time step. 

In the literature, there exist many sensor selection algorithms (see \cite{sp09:boyd} and references therein). In \cite{sp09:boyd}, the sensor selection problem, an integer programming problem, has been relaxed and solved through convex optimization. One popular strategy for sensor selection is to use information driven methods \cite{sp02:Zhao}, \cite{ac10:hoffman}, where the main idea is to select the sensors that provide the most useful information, which is quantified by entropy or mutual information. The posterior Cram\'{e}r-Rao lower bound (PCRLB) is also a very important tool because it provides a theoretical performance limit for a Bayesian estimator. As we have shown in our previous paper \cite{sp10:masazade}, for sensor selection, the complexity to compute the mutual information increases exponentially with the number of sensors to be selected, whereas the computational complexity of Fisher information, which is the inverse of the PCRLB, increases linearly with the number of sensors to be selected. For target tracking in a bearing-only sensor network, a sensor selection approach which minimizes the PCRLB on the estimation error has been proposed in \cite{icassp07:long} and \cite{icassp08:long}, where the selected sensors transmit either analog or quantized data to the fusion center.

For the case where the fusion center receives quantized sensor measurements, given the total bandwidth constraint, $R$, at each time step during tracking, the fusion center should determine the optimal bandwidth distribution for the channels between the sensors and the fusion center which optimizes the target tracking performance in the WSN that consists of $N$ sensors. This problem is more general than the sensor selection problem, because in the bandwidth allocation problem, the channel corresponding to each sensor could be assigned a different number of bits, while in sensor selection problems, a sensor is either activated or not to transmit its measurement under the constraint on the total number of selected sensors. The myopic bandwidth allocation problem can be solved by using an exhaustive search which enumerates all possible bandwidth distributions and decides on the solution that maximizes the determinant of the Fisher information matrix (FIM) which is the inverse of the PCRLB. Under Gaussian assumption, maximizing the determinant of the FIM is equivalent to minimizing the volume of the uncertainty ellipsoid \cite{bar_shalom_tracking_book}. The search space of this problem is $\left(
                                                                  \begin{array}{c}
                                                                    R+N-1 \\
                                                                    N-1 \\
                                                                  \end{array}
                                                                \right)$, which implies that explicit enumeration of all the solutions is computationally prohibitive for large $N$ and $R$. Therefore, computationally efficient suboptimal methods are required.  In \cite{icassp10:onur}, the generalized Breiman, Friedman, Olshen, and Stone (GBFOS) algorithm has been employed for dynamic bandwidth distribution for target tracking which significantly outperforms a static equal bit allocation scheme in terms of tracking performance. But still, as we show later in the paper, the GBFOS algorithm may become computationally costly with increasing values of $N$.

Dynamic programming (DP) \cite{bertsekas2007dynamic} solves the resource allocation problems by breaking them down into simpler steps. For a scalar-valued parameter estimation problem, a DP recursion can be easily formulated to find the optimal bandwidth distribution at each time step by maximizing the Fisher information due to the fact that the total Fisher information is the summation of each sensor's individual Fisher information. For target tracking, even though the Fisher information is in a matrix form and the objective is to maximize the determinant of the FIM, we can still formulate a DP recursion which would yield a suboptimal solution. We refer to this scheme as approximate DP (A-DP), which is computationally very efficient since its complexity increases linearly with $N$.

In our preliminary work \cite{masazade:fusion11}, we compared the performances of dynamic bandwidth allocation approaches based on A-DP, GBFOS and greedy search. Motivated by the sensor selection method presented in \cite{sp09:boyd}, in this paper, we first formulate the bandwidth allocation problem as a constrained optimization problem with binary-valued decision variables and equality constraints.  We then relax and solve the problem optimally using Newton's method by replacing the Boolean variable, $q_{i,m} \in \{0,1\}$, which represents whether or not the quantized measurement of sensor $i$ is transmitted to the fusion center in $m$ bits, with its convex counterpart $\hat{q}_{i,m} \in [0,1]$. Using the idea of probabilistic transmission for bandwidth management \cite{sp09:boyd}, \cite{masazade:ciss10},  we treat $\hat{q}_{i,m}$ as the transmission probability, that is at a given time, sensor $i$ transmits its decision to the fusion center in $m$ bits with probability $\hat{q}_{i,m} \in [0,1]$. Therefore, the convex relaxation based bandwidth allocation method meets the bandwidth constraint in an average sense and introduces a weak constraint on bandwidth availability. We compare the bandwidth allocation schemes based on convex relaxation, A-DP, GBFOS and greedy search in terms of their mean squared error and computational load under different process noise parameters. Simulation results show that convex relaxation, A-DP and GBFOS yield similar tracking performance, which is also similar to that of the optimal bandwidth allocation scheme based on exhaustive search. Among these three suboptimal schemes, A-DP has the least computational load, when the sensor network is large.

The rest of the paper is organized as follows. In Section \ref{section:Target_Tracking}, we introduce the target tracking problem, and describe the optimization of the quantization thresholds and particle filtering in target tracking. In Section \ref{sec:Dynamic_BWA}, we describe the bandwidth distribution schemes based on convex relaxation, A-DP, GBFOS and greedy search. In Section \ref{section:Sim_Results}, we present numerical examples and compare the performances of the considered bandwidth distribution schemes in terms of their computational load and MSEs. Finally, we conclude our work in Section \ref{sec5:Conclusions} and discuss some future research directions.

\section{Target Tracking in Wireless Sensor Networks}
\label{section:Target_Tracking}

The problem we seek to solve is to track a moving target using a WSN where $N$ sensors are grid deployed in a square surveillance area of size $b^2$. The assumption of grid layout is not necessary but has been made here for convenience. Target tracking based on sensor readings can be performed for an arbitrary network layout if sensor placements are known in advance.  All the sensors that are assigned bandwidth report to a central fusion center, which estimates the target state, i.e., the position and the velocity of the target based on quantized sensor measurements.  We assume that the target (e.g., an acoustic or an electromagnetic source) emits a signal from the location ($x_t,y_t$) at time $t$. We assume that the target is based on flat ground and all the sensors and target have the same height so that a 2-D model is sufficient to formulate the problem.

At time $t$, the target dynamics are defined by a 4-dimensional state vector $\mathbf{x}_t = [x_t \quad y_t \quad \dot{x}_t \quad \dot{y}_t]^T$ where $\dot{x}_t$ and $\dot{y}_t$ are the target velocities in the horizontal and the vertical directions respectively. Target motion is defined by the following white noise acceleration model:
\begin{equation}
\label{eq:state_transition_model}
\mathbf{x}_{t+1} = \mathbf{F}\mathbf{x}_t + \mathbf{\upsilon}_t
\end{equation}
where $\mathbf{F}$ models the state dynamics and $\mathbf{\upsilon}_t$ is the process noise which is assumed to be
white, zero-mean and Gaussian with the following covariance matrix $\mathbf{Q}$.
\begin{equation}
\label{eq:Necessary_matrices}
\mathbf{F} = \left[
  \begin{array}{cccc}
    1 & 0 & {\cal D} & 0 \\
    0 & 1 & 0 & {\cal D} \\
    0 & 0 & 1 & 0 \\
    0 & 0 & 0 & 1 \\
  \end{array}
\right] \:,\: \mathbf{Q} = \rho \left[
  \begin{array}{cccc}
    \frac{{\cal D}^3}{3} & 0 & \frac{{\cal D}^2}{2} & 0 \\
    0 & \frac{{\cal D}^3}{3} & 0 & \frac{{\cal D}^2}{2} \\
    \frac{{\cal D}^2}{2} & 0 & {\cal D} & 0 \\
    0 & \frac{{\cal D}^2}{2} & 0 & {\cal D} \\
  \end{array}
\right]
\end{equation}
In (\ref{eq:Necessary_matrices}), ${\cal D}$ and $\rho$ denote the time interval between adjacent sensor measurements and the process noise parameter, respectively. It is assumed that the fusion center has perfect information about the target state-space model (\ref{eq:state_transition_model}) as well as the process noise statistics (\ref{eq:Necessary_matrices}).

The target is assumed to be an acoustic or an electromagnetic source that follows the power attenuation model provided
below \cite{niu2005distributed}.  At any given time $t$, the signal power received at the sensor $i$ is given as
\begin{equation}
\label{eq:power_attenuation}
a_{i,t}^2 = \frac{P_0}{1+\alpha d_{i,t}^n}
\end{equation}
By adopting this model, we prevent the receiver amplifier from saturation and the regularity conditions for PCRLB hold when the target is very close to a sensor. In Eq. (\ref{eq:power_attenuation}), $P_0$ denotes the signal power of the target, $n$ is the signal decay exponent and $\alpha$ is a scaling parameter. $d_{i,t}$ is the distance between the target and the $i^{th}$ sensor, $d_{i,t} = \sqrt{(x_i - x_t)^2 + (y_i - y_t)^2}$, where $(x_i, y_i)$ are the coordinates of the $i^{th}$ sensor. Without loss of generality, $\alpha$ and $n$ are assumed to be unity and 2, respectively. At time $t$, the received signal at sensor $i$ is given by
\begin{equation}
z_{i,t} = a_{i,t} + n_{i,t}
\end{equation}
where $n_{i,t}$ is the noise term modeled as additive white Gaussian noise (AWGN), i.e., $n_{i,t} \sim
{\cal N}(0, \sigma^2)$, which represents the cumulative effects of sensor background noise and the modeling error
of signal parameters.

Rather than transmitting analog sensor observations to the fusion center, transmitting a quantized version of sensor
measurements decreases the amount of communication and therefore reduces the energy consumption. A sensor measurement
$z_{i,t}$ at sensor $i$ is locally quantized before its transmission to the fusion center using $R_{i,t}$ bits. Let $\mathbf{R}_t \triangleq [R_{1,t},\ldots,R_{N,t}]$ be the vector of quantization rates used by the $N$ sensors in the network. For the bandwidth distribution problem, at each time step of tracking, $R_{i,t}$ can take values $m, \:\: m \in \{0,1,\ldots,R\}$ where $R$ is the maximum number of bits to be transmitted to the fusion center collectively by all the sensors. Let $L_m \triangleq 2^{m}-1$ be the number of decision intervals for transmitting $m$ bits to the fusion center and $D_{i,t}$ be the $m$-bit observation of sensor $i$ quantized with rate $R_{i,t} = m$ at time step $t$, then
\begin{equation}
D_{i,t} =  \left\{ \begin{array}{rl}
    \label{eq:quant_rule}
  0 &\mbox{ $-\infty < z_{i,t} <\eta^{m}_{1}$} \\
  1 &\mbox{ $\eta^{m}_{1} < z_{i,t} <\eta^{m}_{2}$}  \\
\vdots\\
  L_{m}-1 &\mbox{ $\eta^{m}_{L_{m}-1} < z_{i,t} <\infty$ }  \\
       \end{array} \right.
\end{equation} where $\boldsymbol{\eta}^{m} = [\eta^{m}_{0}\; \eta^{m}_{1}\; \ldots \; \eta^{m}_{L_{m}}]$ with $\eta^{m}_{0} = -\infty$ and $\eta^{m}_{L_m} = \infty$. The quantization thresholds are assumed to be identical at each sensor for simplicity. We explain the selection of the quantization thresholds for each data rate $R_{i,t} = m$ later in this section. Given $\mathbf{x}_t$ and $m$, it is easy to show that the probability of a particular quantization output $l$ is,
\begin{equation}
\label{eq:ML_Decision}
P(D_{i,t}=l|\mathbf{x}_t, R_{i,t} = m) = Q\left(\frac{\eta^{m}_{l}-a_{i,t}}{\sigma}\right)-Q\left(\frac{\eta^{m}_{l+1}-a_{i,t}}{\sigma}\right)
\end{equation} where $Q(.)$ is the complementary distribution function of the standard Gaussian distribution with zero mean and unit variance,
\begin{equation}
Q(x) =\int_x^{\infty} \frac{1}{\sqrt{2\pi}}\exp\left(-\frac{t^2}{2}\right)dt
\end{equation}
At time $t$, let the fusion center receive the data vector $\mathbf{D}_t = [D_{1,t},\ldots,D_{N,t}]$ from the $N$ sensors with the corresponding quantization rate vector $\mathbf{R}_t = [R_{1,t},\ldots,R_{N,t}]$, then
\begin{equation}
\label{eq:ML_Decision_all}
p(\mathbf{D}_{t}|\mathbf{x}_{t}, \mathbf{R}_t) = \prod_{i=1}^N p(D_{i,t}|\mathbf{x}_{t}, R_{i,t})
\end{equation}
where we assume $p(D_{i,t}|\mathbf{x}_{t}, R_{i,t}=0) = 1$.

\subsection{PCRLB with quantized data}

Let $p(\mathbf{D}_t,\mathbf{x}_t)$ be the joint probability density of $\mathbf{D}_t$ and $\mathbf{x}_t$, and $\hat{\mathbf{x}}_t$ be an estimate of $\mathbf{x}_t$ at time step $t$. Based on the received data $\mathbf{D}_t$ quantized with rate vector $\mathbf{R}_t$, and the prior probability distribution function of $\mathbf{x}_t$, $p(\mathbf{x}_t)$, the PCRLB on the mean squared estimation error has the form,
\begin{equation}
E\left\{[\hat{\mathbf{x}}_t-\mathbf{x}_t][\hat{\mathbf{x}}_t-\mathbf{x}_t ]^T  | \mathbf{R}_t \right\} \geq \mathbf{J_t^{-1}(\mathbf{R}_t)}
\end{equation}
where $\mathbf{J}_t(\mathbf{R}_t)$ is the $4 \times 4$ Fisher information matrix (FIM) with the elements
\begin{equation}
\label{eq:Fisher_partial}
\mathbf{J}_t(\mathbf{R}_t)(i,j) = E\left[ -\frac{\partial^2 \log{p(\mathbf{D}_t,\mathbf{x}_t | \mathbf{R}_t)}}{\partial \mathbf{x}_t(i) \partial \mathbf{x}_t(j)} \right] \quad i,j \in \{1,\ldots,4\}
\end{equation} where $\mathbf{J}_{t} (\mathbf{R}_{t}) (i,j)$ denotes the $i^{th}$ row, $j^{th}$ column element of the matrix $\mathbf{J}_{t} (\mathbf{R}_{t})$ and $\mathbf{x}_t(i)$ denotes the $i^{th}$ element of vector $\mathbf{x}_t$. Let $\nabla_{\mathbf{x}_t}^{\mathbf{x}_t} \triangleq \nabla_{\mathbf{x}_t}\nabla_{\mathbf{x}_t}^T$ denote the
second order partial derivative operator with respect to $\mathbf{x}_t$. Using this notation, (\ref{eq:Fisher_partial}) can be rewritten in a more compact fashion as,
\begin{equation}
\label{eq:Fisher_partial2}
\mathbf{J}_t (\mathbf{R}_t) = E\left[ -\nabla_{\mathbf{x}_t}^{\mathbf{x}_t} \log{p(\mathbf{D}_t,\mathbf{x}_t | \mathbf{R}_t)}\right]
\end{equation} Since $p(\mathbf{D}_t,\mathbf{x}_t|\mathbf{R}_t) = p(\mathbf{D}_t|\mathbf{x}_t, \mathbf{R}_t)p(\mathbf{x}_t)$, $\mathbf{J}_t (\mathbf{R}_t)$ can be decomposed into two parts as,
\begin{eqnarray}
\label{eq:J_D_J_P}
&& \mathbf{J}_t (\mathbf{R}_t) = \mathbf{J}_t^D (\mathbf{R}_t) + \mathbf{J}_t^P
\end{eqnarray} where
\begin{eqnarray}
&& \mathbf{J}_t^D (\mathbf{R}_t) \triangleq  E_{p(\mathbf{D}_t|\mathbf{x}_t)p(\mathbf{x}_t)}\left[ -\nabla_{\mathbf{x}_t}^{\mathbf{x}_t} \log{p(\mathbf{D}_t|\mathbf{x}_t, \mathbf{R}_t)}\right] \nonumber \\
&& \mathbf{J}_t^P \triangleq  E_{p(\mathbf{x}_t)}\left[ -\nabla_{\mathbf{x}_t}^{\mathbf{x}_t} \log{p(\mathbf{x}_t)}\right] \nonumber
\end{eqnarray}
Note that $\mathbf{J}_t^D (\mathbf{R}_t)$ represents the Fisher information obtained from the data averaged over the prior distribution $p(\mathbf{x}_t)$ and $\mathbf{J}_t^P$ represents the \textit{a priori} Fisher information. $E_{p(\mathbf{D}_t|\mathbf{x}_t)p(\mathbf{x}_t)}[.]$ and $E_{p(\mathbf{x}_t)}[.]$ denote expectations with respect to $p(\mathbf{D}_t|\mathbf{x}_t)p(\mathbf{x}_t)$ and $p(\mathbf{x}_t)$ respectively.

Given the vector of quantization rates $\mathbf{R}_t = [R_{1,t},\ldots,R_{N,t}]$ and using (\ref{eq:ML_Decision_all}) in (\ref{eq:J_D_J_P}), the data part of the Fisher information can be written as,
\begin{eqnarray}
\label{eq:J_D_uzun}
&\mathbf{J}_t^D(R_{1,t},\ldots,R_{N,t}) &  = \int_{\mathbf{x}_t} E_{p(\mathbf{D}_t|\mathbf{x}_t,\mathbf{R}_t)} \left[ -\nabla_{\mathbf{x}_t}^{\mathbf{x}_t} \log{p(\mathbf{D}_t|\mathbf{x}_t, \mathbf{R}_t)} \right]  p(\mathbf{x}_t) d \mathbf{x}_t  \\
&& = \sum_{i=1}^N \int_{\mathbf{x}_t}  \left\{ \sum_{l=0}^{2^{R_{i,t}-1}} -\nabla_{\mathbf{x}_t}^{\mathbf{x}_t} \log{p({D}_{i,t} = l|\mathbf{x}_t, {R}_{i,t})} p({D}_{i,t} = l |\mathbf{x}_t, {R}_{i,t}) \right\} p(\mathbf{x}_t) d \mathbf{x}_t \nonumber
\end{eqnarray}
For a given $\mathbf{x}_t$, let us define $\mathbf{J}_{i,t}^S(R_{i,t}|\mathbf{x}_t)$, as the Fisher information of sensor $i$,
\begin{eqnarray}
\label{eq:J_S_uzun}
& \mathbf{J}_{i,t}^S(R_{i,t}|\mathbf{x}_t)&  \triangleq  E_{p(\mathbf{D}_t|\mathbf{x}_t,\mathbf{R}_t)}\left[-\nabla_{\mathbf{x}_t}^{\mathbf{x}_t} \log p(D_{i,t}|\mathbf{x}_t)\right]  \\
& &  =  \sum_{l=0}^{2^{R_{i,t}}-1} \left\{-\nabla_{\mathbf{x}_t}^{\mathbf{x}_t} \log{p({D}_{i,t} = l|\mathbf{x}_t, {R}_{i,t})} p({D}_{i,t} = l|\mathbf{x}_t, {R}_{i,t})  \right\} \nonumber
\end{eqnarray}
Then combining (\ref{eq:J_D_uzun}) and (\ref{eq:J_S_uzun}), sensor $i$'s contribution to the Fisher information $\mathbf{J}_{i,t}^D(R_{i,t})$ can be stated as,
\begin{eqnarray}
\label{eq:sensor_data_FI}
& \mathbf{J}_{i,t}^D(R_{i,t}) \triangleq &  \int_{\mathbf{x}_t}  \mathbf{J}_{i,t}^S(R_{i,t}|\mathbf{x}_t) p(\mathbf{x}_t) d \mathbf{x}_t
\end{eqnarray}
Given $\mathbf{R}_t$, the Fisher information at time $t$ can be written as,
\begin{equation}
\label{eq:Fisher_Summable}
\mathbf{J}_t (\mathbf{R}_t) = \sum_{i=1}^N \mathbf{J}_{i,t}^D(R_{i,t}) + \mathbf{J}_t^P
\end{equation} From (\ref{eq:J_S_uzun}), after straight-forward calculations, the $(1,1)$ term of $\mathbf{J}_{i,t}^S(R_{i,t}|\mathbf{x}_t)$ can be derived as,
\begin{eqnarray}
\label{eq:app_lazim}
E \left[-\frac{\partial^2 \log p(D_{i,t}|\mathbf{x}_t)}{\partial x_t^2} \right]= \sum_{l=0}^{2^{R_{i,t}}-1} \frac{1}{p(D_{i,t} = l|\mathbf{x}_t,R_{i,t})} \left(\frac{\partial p(D_{i,t} = l|\mathbf{x}_t,R_{i,t})}{\partial x_t}\right)^2
\end{eqnarray} The rest of the terms can be derived similarly. Using the procedures similar to \cite{NiuVarshney:sp06}, $\mathbf{J}_{i,t}^S(R_{i,t}|\mathbf{x}_t)$ can be obtained as follows,
\begin{eqnarray}
\label{eq:mle_fim2}
&& \mathbf{J}_{i,t}^S(R_{i,t} = m | \mathbf{x}_t) = n^2 \kappa_{i,t} (m,x_i,y_i,x_t,y_t) \frac{a_{i,t}^2 \alpha^2 d_{i,t}^{2n-4}}{(1+\alpha d_{i,t}^n)^2} \times \\
&&  \left[ \begin{array}{cccc}
(x_i-x_t)^2  & (x_i-x_t)(y_i-y_t) & 0 & 0\\
(x_i-x_t)(y_i-y_t)   & (y_i-y_t)^2 & 0 & 0\\
0  & 0 & 0 & 0\\
0  & 0 & 0 & 0\\
       \end{array} \right] \nonumber
\end{eqnarray}
where
\begin{eqnarray}
\label{eq:kappa}
&& \kappa_{i,t}(m,x_i,y_i,x_t,y_t) =  \frac{1}{8\pi\sigma^2} \left\{ \sum_{l=0}^{2^{m}-1}\frac{\left[ e^{-\frac{{(\eta^m_{l}-a_{i,t})}^2}{2\sigma^2}}-e^{-\frac{{(\eta^m_{l+1}-a_{i,t})}^2}{2\sigma^2}}\right]^2} {p(D_i=l |\mathbf{x}_t)} \right\}
\end{eqnarray} Detailed derivation of (18) can be found in the Appendix. Note that in (\ref{eq:mle_fim2}) and (\ref{eq:kappa}), $d_{i,t}$ and $a_{i,t}$ are functions of the sensor location $(x_i,y_i)$ and target location $(x_t,y_t)$.

\subsection{Optimization of Quantization Thresholds}

The Fisher information and hence the PCRLB are functions of the quantization thresholds corresponding to each data rate $R_{i,t} = m$. Thus, the quantization thresholds should be designed to achieve better estimation accuracy. An algorithm to obtain the optimal quantization thresholds that minimizes the variance of the estimation errors has been proposed in \cite{NiuVarshney:sp06}. If we assume that $(x_i,y_i)$ and $(x_t,y_t)$ are uniformly distributed in a region, we can minimize the sum of two diagonal elements of the CRLB matrix, after averaging the CRLB matrix over all the random parameters which may result in a large computational load since it requires a multiple fold integration. To alleviate this problem,  some alternative methods to design the quantization thresholds were developed in \cite{NiuVarshney:sp06}.

Note that all the information about $[x_t,y_t]^T$ is contained in sensors' signal amplitudes $(a_{i,t})$'s. If all the signal amplitudes can be recovered from their quantized data $D_{i,t}$ accurately, an accurate estimate of $[x_t,y_t]^T$ can be obtained. In this paper, we use the Fisher information based heuristic quantization method \cite{NiuVarshney:sp06} which maximizes the Fisher information about the signal amplitude $a_{i,t}$ contained in the quantized data $D_{i,t}$. We define $F_a (\boldsymbol{\eta}|x_i, y_i,x_t, y_t,R_{i,t}=m)$ as the Fisher information of the signal amplitude contained in quantized $m$-bit data, $D_{i,t}$, using a threshold $\boldsymbol{\eta}$. Note that $a_{i,t}$ is a function of $d_{i,t}$ for fixed $P_0$, $\alpha$ and $n$ as defined in (\ref{eq:power_attenuation}). Then given $R_{i,t} = m$, sensor location $(x_i, y_i)$ and source location $(x_t, y_t)$, it has been derived in \cite{NiuVarshney:sp06} that $F_a (\boldsymbol{\eta}|x_i, y_i,x_t, y_t,R_{i,t}=m) = 4\kappa_{i,t}(m,x_i, y_i,x_t, y_t)$. The Fisher information based heuristic quantization method \cite{NiuVarshney:sp06} finds the decision thresholds that maximize
\begin{eqnarray}
& F_a (\boldsymbol{\eta}|R_{i,t}=m) & = E[-\nabla_{a_{i,t}}^{a_{i,t}} \log p(D_{i,t}|a_{i,t}(x_i, y_i,x_t, y_t))] \\
&& = \int_{x_i, y_i,x_t, y_t} 4 \kappa(m,x_i, y_i,x_t, y_t,)dx_i dy_i dx_t d y_t \nonumber \\
&& = \int_{u} 4 \kappa(m|u)p(u)du \nonumber
\end{eqnarray}
where $u = d_{i,t}^2$ and the Fisher information about the signal amplitude is averaged over the probability density function of $u$, $p(u)$, under the assumption that $(x_i,y_i)$ and $(x_t,y_t)$ are independent and identically distributed and follow a uniform distribution $U[-b/2, b/2]$. Derivation of $p(u)$ and other details of this quantizer design approach can be found in \cite{NiuVarshney:sp06}. We assume that the decision thresholds of each quantization rate are identical at each sensor. The quantization thresholds of each possible quantization rate is optimized offline and can be stored at each sensor before the WSN operation.

\subsection{Particle Filtering with Quantized Data}

It is known that Kalman Filter provides the optimal solution to the Bayesian sequential estimation problem for linear and Gaussian systems. In nonlinear systems, the extended Kalman filter (EKF) can be used to provide a suboptimal
solution by linearizing the nonlinear state dynamics and/or nonlinear measurement equations locally. However, it has been shown \cite{Willet:aes08} that, even for linear and Gaussian systems, when the sensor measurements are quantized, the EKF fails to provide an acceptable performance especially when the number of quantization levels is small. Therefore, we propose to employ a particle filter to solve the Bayesian sequential estimation problem.

Let $\mathbf{D}_{1:t} = [\mathbf{D}_{1},\ldots, \mathbf{D}_{t}]$ be the received sensor data up to time $t$ which are obtained according to the data rates $\mathbf{R}_{1:t} = [\mathbf{R}_{1},\ldots, \mathbf{R}_{t}]$. In particle filtering, the main idea is to find a discrete representation of the posterior distribution $p(\mathbf{x}_{t}|\mathbf{D}_{1:t})$ by using a set of particles $\{\mathbf{x}_t^s;\: s = 1,\ldots,N_s\}$ with associated weights $\{w_t^s;\: s = 1,\ldots,N_s\}$. The posterior density at $t$ can be approximated as,
\begin{equation}
\label{eq:particle_filter_original}
p(\mathbf{x}_{t}|\mathbf{D}_{1:t}) \approx \sum_{s=1}^{N_s} w_t^s \delta(\mathbf{x}_{t} - \mathbf{x}_{t}^s)
\end{equation} where $N_s$ denotes the total number of particles. In this paper, we employ sequential importance resampling (SIR) particle filtering algorithm \cite{Particle_tutorial} to solve the nonlinear Bayesian filtering problem. In Algorithm \ref{alg1}, we provide a summary of the SIR based particle filtering rather than discussing the details. Note that $T_S$ in Algorithm \ref{alg1} denotes the number of time steps over which the target is tracked.  A more detailed treatment of particle filtering can be found in a wide variety of publications such as \cite{Particle_tutorial}.

\begin{algorithm}                      
\caption{SIR based Particle Filtering for Target Tracking}          
\label{alg1}                           
\begin{algorithmic}                    
\STATE Set $t = 0$. Generate initial particles $\mathbf{x}_{0}^{s} \sim p(\mathbf{x}_0)$ with $\forall s\;, w_0^s = N_s^{-1}$.
\WHILE {$t \leq T_S$}
\STATE (A1.1) $\mathbf{x}_{t+1}^{s} = \mathbf{F}\mathbf{x}_{t}^{s} +  \mathbf{\upsilon}_t$ (Propagating particles)
\STATE (A1.2) $p(\mathbf{x}_{t+1}|\mathbf{D}_{1:t}) = \frac{1}{N_s}\sum_{s=1}^{N_s} \delta(\mathbf{x}_{t+1} - \mathbf{x}_{t+1}^{s}) $
\STATE (A1.3) Bandwidth Allocation: Decide $\mathbf{R}_{t+1}$ and obtain sensor data $\mathbf{D}_{t+1}$
\STATE (A1.4) $w_{t+1}^{s} \propto  p(\mathbf{D}_{t+1}|\mathbf{x}_{t+1}^{s}, \mathbf{R}_{t+1})$ (Updating weights)
\STATE  $w_{t+1}^{s} = \frac{w_{t+1}^{s}}{\sum_{j=1}^{N_s} w_{t+1}^{j}}$ (Normalizing weights)
\STATE  $\mathbf{\hat{x}}_{t+1} = \sum_{s=1}^{N_s} w_{t+1}^{s} \mathbf{x}_{t+1}^{s}$
\STATE (A1.5) $\{\mathbf{x}_{t+1}^{s},N_s^{-1}\} = \textrm{Resampling} (\mathbf{x}_{t+1}^{s},w_{t+1}^{s}) $
\STATE (A1.6) $t = t+1$
  \ENDWHILE
\end{algorithmic}
\end{algorithm} In Algorithm \ref{alg1}, $p(\mathbf{D}_{t+1}|\mathbf{x}_{t+1}^{s}, \mathbf{R}_{t+1})$ is obtained according to (\ref{eq:ML_Decision}) and (\ref{eq:ML_Decision_all}). Resampling step avoids the situation that all but one of the importance weights are close to zero \cite{Particle_tutorial}.

By using equations (\ref{eq:Fisher_partial}) to (\ref{eq:Fisher_Summable}), at time $t$, one can compute the PCRLB on the estimation error and the corresponding FIM, for a given bandwidth allocation scheme $\mathbf{R}_t$ and prior distribution $p(\mathbf{x}_t)$. For the bandwidth allocation problem, at time $t$, from (A1.2), we first generate the prior $p(\mathbf{x}_{t+1}|\mathbf{D}_{1:t})$ using data received up to time $t$.

Under the Gaussian assumption, maximizing the determinant of the FIM is equivalent to minimizing the volume of the uncertainty ellipsoid \cite{bar_shalom_tracking_book}. Therefore, we determine bandwidth allocation scheme for time $t+1$, $\mathbf{R}_{t+1}$, by maximizing the determinant of the Fisher information about $\mathbf{x}_{t+1}$ as,
\begin{eqnarray}
\label{eq:maxDetFIM}
&\max_{R_{1,t+1},\ldots,R_{N,t+1}} &  \det(\mathbf{J}_{t+1}(\mathbf{R}_{t+1})) \\
&\mathrm{s.t.} &  \sum_{i=1}^N R_{i,t+1} = R \nonumber
\end{eqnarray}
The fusion center then informs the sensors about $\mathbf{R}_{t+1}$ and sensors transmit their quantized measurements $\mathbf{D}_{t+1}$ accordingly. The Fisher information, $\mathbf{J}_{t+1}(\mathbf{R}_{t+1})$ is written as
\begin{eqnarray}
\mathbf{J}_{t+1} (\mathbf{R}_{t+1}) =  E_{p(\mathbf{D}_{t+1},\mathbf{x}_{t+1}|\mathbf{D}_{1:t},\mathbf{R}_{t+1})}  \left[-\nabla_{\mathbf{x}_{t+1}}^{\mathbf{x}_{t+1}} \log p(\mathbf{D}_{t+1},\mathbf{x}_{t+1}|\mathbf{D}_{1:t},\mathbf{R}_{t+1}) )  \right] \nonumber
\end{eqnarray}
Following the derivations from (\ref{eq:Fisher_partial}) to (\ref{eq:Fisher_Summable}), the Fisher information, $\mathbf{J}_{t+1} (\mathbf{R}_{t+1})$,  is obtained as,
\begin{equation}
\label{eq:FIM_BA}
\mathbf{J}_{t+1} (\mathbf{R}_{t+1}) = \sum_{i=1}^N \mathbf{J}_{t+1}^D(R_{i,t+1}) + \mathbf{J}_{t+1}^P
\end{equation} Using the particle approximation,
\begin{equation}
\label{eq:particle_filter_approx}
p(\mathbf{x}_{t+1}|\mathbf{D}_{1:t}) \approx \frac{1}{N_S} \sum_{s=1}^{N_S}  \delta(\mathbf{x}_{t+1}-\mathbf{x}_{t+1}^s) \end{equation}
$\mathbf{J}_{t+1}^D(R_{i,t+1})$ is found from,
\begin{eqnarray}
\mathbf{J}_{t+1}^D(R_{i,t+1}) = \frac{1}{N_S}\sum_{s=1}^{N_S} \mathbf{J}_{t+1}^S(R_{i,t+1}|\mathbf{x}_{t+1}^s)
\end{eqnarray}
As in (\ref{eq:J_D_J_P}), $\mathbf{J}_{t+1}^P = E_{p(\mathbf{x}_{t+1}|\mathbf{D}_{1:t})}[-\nabla_{\mathbf{x}_{t+1}}^{\mathbf{x}_{t+1}} \log p(\mathbf{x}_{t+1}|\mathbf{D}_{1:t})]$ has been defined as the prior Fisher information of
$\mathbf{x}_{t+1}$. According to (\ref{eq:particle_filter_approx}), $p(\mathbf{x}_{t+1}|\mathbf{D}_{1:t})$ has a non-parametric representation by a set of random particles with associated weights, so it is very difficult to calculate the exact $\mathbf{J}_{t+1}^P$ \cite{icassp11:onur}. Instead, we use a Gaussian approximation such that
$p(\mathbf{x}_{t+1}|\mathbf{D}_{1:t}) \approx {\cal N} (\boldsymbol{\mu}_{t+1}, \boldsymbol{\Sigma}_{t+1})$, where
\begin{eqnarray}
&& \boldsymbol{\mu}_{t+1} = \frac{1}{N_s} \sum_{s=1}^{N_s} \mathbf{x}_{t+1}^s \nonumber
\end{eqnarray} and
\begin{eqnarray}
&& \boldsymbol{\Sigma}_{t+1} = \frac{1}{N_s} \sum_{s=1}^{N_s} (\mathbf{x}_{t+1}^s-\boldsymbol{\mu}_{t+1})(\mathbf{x}_{t+1}^s-\boldsymbol{\mu}_{t+1})^T \nonumber
\end{eqnarray}
Given the Gaussian approximation, it is easy to show that $\mathbf{J}_{t+1}^P = \boldsymbol{\Sigma}_{t+1}^{-1}$.

\section{Dynamic Bandwidth Allocation for Target Tracking}
\label{sec:Dynamic_BWA}

An exhaustive search can be employed to find the optimal bandwidth distribution which maximizes (\ref{eq:maxDetFIM}). For a network of $N$ sensors and bandwidth constraint $R$, there are a total of $\left(
                                                                          \begin{array}{c}
                                                                            R+N-1 \\
                                                                            N-1 \\
                                                                          \end{array}
                                                                        \right) = \frac{(N+R-1)!}{(N-1)!R!}
$ possible bandwidth distribution solutions. For large $N$ and $R$, such an exhaustive search may not be
feasible in real time. Therefore suboptimal but computationally more efficient algorithms are required which we explore in this section.

\subsection{Convex Optimization Based Dynamic Bandwidth Allocation}

In this paper, we use the log determinant of the FIM as the objective function for resource management.  Using Boolean variables $q_{i,m} \in \{0,1\}$, the bandwidth allocation problem can be explicitly formulated as follows,
\begin{eqnarray}
\label{eq:FIM_CnvX_boolean}
& \max_{\textbf{q}_{t+1}} & \log\det(\mathbf{J}_{t+1}(\textbf{q}_{t+1}))  = \log\det\left(\sum_{m=0}^R \sum_{i=1}^N q_{i,m}\mathbf{J}_{i,t+1}^D(R_{i,t+1} = m) + \mathbf{J}_{t+1}^P \right)\\
& \textrm{subject to} & \sum_{m=0}^R q_{i,m} = 1 \quad i \in \{1,\ldots,N\} \nonumber \\
&& \sum_{m=0}^R \sum_{i=1}^N m\: q_{i,m} = R \nonumber \\
&& q_{i,m} \in \{0,1\} \quad m \in \{0,1,\ldots,R\} \quad i \in \{1,\ldots,N\} \nonumber
\end{eqnarray} In the above formulation, $\textbf{q}_{t+1} = [q_{1,0}, q_{2,0},\ldots,q_{N,0},\ldots,q_{1,R}, q_{2,R},\ldots,q_{N,R}]^T$ denotes the bandwidth allocation scheme for time $t+1$ where $q_{i,m} = 1$ when sensor $i$ transmits its measurement in $m$ bits and $\mathbf{J}_{i,t+1}^D(m)$ is the corresponding FIM of sensor $i$. Note that we drop the time index $t+1$ from the elements of vector $\mathbf{q}_{t+1}$ to simplify the notation. All constraints are equality constraints where the first $N$ constraints guarantee that each sensor can transmit using only one of the quantization rates. If $m=0$ is selected, the quantized measurement of the sensor is not transmitted to the fusion center. The $(N+1)^{th}$ constraint ensures that the sum rate constraint is satisfied and the last $N(R+1)$ constraints restrict $q_{i,m}$ to be Boolean.

Similar to the convex relaxation approach presented in \cite{sp09:boyd}, the last $N(R+1)$ constraints can be relaxed by replacing the Boolean variables $q_{i,m} \in \{0,1\}$ with their continuous counterparts,  $\hat{q}_{i,m} \in [0,1]$. Then the problem becomes
\begin{eqnarray}
\label{eq:FIM_CnvX_relaxed}
& \max_{\mathbf{\hat{q}}_{t+1}} & \log\det(\mathbf{J}_{t+1}(\mathbf{\hat{q}}_{t+1}))  = \log\det\left(\sum_{m=0}^R \sum_{i=1}^N \hat{q}_{i,m}\mathbf{J}_{i,t+1}^D(R_{i,t+1} = m) + \mathbf{J}_{t+1}^P \right)\\
& \textrm{subject to} & \sum_{m=0}^R \hat{q}_{i,m} = 1 \quad i \in \{1,\ldots,N\} \nonumber \\
&& \sum_{m=0}^R \sum_{i=1}^N m\: \hat{q}_{i,m} = R \nonumber \\
&& 0 \leq \hat{q}_{i,m} \leq 1 \quad m \in \{0,1,\ldots,R\} \quad i \in \{1,\ldots,N\} \nonumber
\end{eqnarray}
We can further relax the problem (\ref{eq:FIM_CnvX_relaxed}), by removing the last $N(R+1)$ constraints and including them into the objective function. Then, the new cost function to be minimized becomes as,
\begin{eqnarray}
&& \phi(\mathbf{\hat{q}}_{t+1}) \triangleq \\
&&  -\left\{\log{\det{\left(\sum_{m=0}^R \sum_{i=1}^N \hat{q}_{i,m} \mathbf{J}_{i,t+1}^D(m) + \mathbf{J}_{t+1}^P \right)}} + \tau \sum_{m=0}^R \sum_{i=1}^N \Big(\log{(\hat{q}_{i,m})} + \log{(1-\hat{q}_{i,m})}\Big) \right\}\nonumber
\end{eqnarray}
where $\phi(\mathbf{\hat{q}}_{t+1})$ is a convex function of the decision variables $\hat{q}_{i,m}$ \cite{boyd2004convex}. The additional summation term in the objective function forces $\hat{q}_{i,m}$ to be in the interval $[0,1]$. $\tau$ is a positive parameter that controls the quality of the approximation.

Let us define,
\begin{eqnarray}
&& {\cal A} \triangleq \left(
  \begin{array}{ccccccccccccc}
    1 & 0 & \ldots & 0 &  1 & 0 & \ldots & 0 & \ldots & 1 & \ldots & 0 & 0 \\
    0 & 1 & \ldots & 0 &  0 & 1 & \ldots & 0 & \ldots &  0 & \ldots & 0 & 0 \\
    : & : & : & : & : &  : & : & : & : & : & : & : & :\\
    0 & 0 & \ldots & 1 &  0 & 0 & \ldots & 1 &  \ldots & 0 & \ldots & 0 & 1 \\
    0 & 0 & \ldots & 0 &  1 & 1 & \ldots & 1 &  \ldots & R & \ldots & R & R \\
  \end{array}
\right) \nonumber \\
&& \mathbf{\hat{q}}_{t+1} \triangleq \left( \begin{array}{ccccccccccccc}
          \hat{q}_{1,0} & \hat{q}_{2,0} & \dots & \hat{q}_{N,0} & \hat{q}_{1,1} & \hat{q}_{2,1} & \ldots & \hat{q}_{N,1} & \ldots  & \hat{q}_{1,R} & \ldots & \hat{q}_{N-1,R} & \hat{q}_{N,R} \\
        \end{array} \right)^T \nonumber \\
&& \mathrm{and}     \nonumber   \\
&& \mathbf{b}  \triangleq \left( \begin{array}{ccccc}
          1 & 1 & \ldots & 1 & R \\
        \end{array} \right)^T \nonumber
\end{eqnarray}
Then, the first $N+1$ equality constraints of (\ref{eq:FIM_CnvX_relaxed}) can be represented in a matrix form as,
\begin{eqnarray}
&& {\cal A} \mathbf{\hat{q}}_{t+1} = \mathbf{b} \nonumber
\end{eqnarray}
Finally, we have the following convex optimization problem,
\begin{eqnarray}
\label{eq:Prob_BA}
& \min_{\mathbf{\hat{q}}_{t+1}} & \phi (\mathbf{\hat{q}}_{t+1}) \\
& \textrm{subject to} & {\cal A} \mathbf{\hat{q}}_{t+1} = \mathbf{b} \nonumber
\end{eqnarray} which can be solved efficiently and optimally using Newton's method. The underdetermined system ${\cal A}\mathbf{\hat{q}}_{t+1} = \mathbf{b}$ has infinite number of solutions but there is only a subset of solutions which are feasible satisfying $\mathbf{0} \leq \mathbf{\hat{q}}_{t+1} \leq \mathbf{1}$ where $\mathbf{0}$ and $\mathbf{1}$ are the all zero and all one vectors respectively. The Newton method starts with a feasible solution, so we formulate the following linear optimization sub-problem to find an initial feasible solution,
\begin{eqnarray}
\label{eq:linprog}
& \min_{\mathbf{\hat{q}}_{t+1}} & -\sum_{m=0}^R \sum_{i=1}^N \hat{q}_{i,m} \\
& \textrm{subject to} & {\cal A}\mathbf{\hat{q}}_{t+1} = \mathbf{b} \nonumber \\
&& 0 \leq \hat{q}_{i,m} \leq 1 \quad m \in \{0,\ldots,R\} \quad i \in \{1,\ldots,N\} \nonumber
\end{eqnarray}
The optimality conditions for (\ref{eq:Prob_BA}), which is named as the Karush Kuhn Tucker (KKT) system, is written as  \cite{boyd2004convex},
\begin{equation}
\label{eq:Newton_KKT}
\left(
  \begin{array}{cc}
    \nabla_{\hat{q}_{t+1}}^{\hat{q}_{t+1}} \phi  & {\cal A}^T \\
    {\cal A} & 0 \\
  \end{array}
\right) \left( \begin{array}{c}
          \Delta \mathbf{Z} \\
          \omega
        \end{array} \right) = \left( \begin{array}{c}
          -\nabla_{\hat{q}_{t+1}} \phi \\
          0
        \end{array} \right)
\end{equation}
where $\Delta \mathbf{Z}$ is the Newton Step, $\omega$ is the the optimal dual variable, $\nabla_{\hat{q}_{t+1}} \phi$ is the gradient vector, and $\nabla_{\hat{q}_{t+1}}^{\hat{q}_{t+1}} \phi$ is the Hessian matrix of $\phi$ with respect to the decision vector $\hat{\mathbf{q}}_{t+1}$.

In order to solve the system given in (\ref{eq:Newton_KKT}), first we need to compute the gradient vector, $\nabla_{\hat{q}_{t+1}} \phi$, and the Hessian matrix, $\nabla_{\hat{q}_{t+1}}^{\hat{q}_{t+1}} \phi$, with sizes $N(R+1) \times 1$ and $N(R+1) \times N(R+1)$ respectively. Let us start by computing the gradient.  First, the $(i,m)^{th}$ element of the gradient vector is,
\begin{eqnarray}
&& (\nabla_{\hat{q}_{t+1}} \phi)_{i,m} = \\
&   &  \frac{\partial}{\partial \hat{q}_{i,m}}  \left\{-\log{\det{\left(\sum_{m=0}^R \sum_{i=1}^N \hat{q}_{i,m} \mathbf{J}_{i,t+1}^D (m) + \mathbf{J}_{t+1}^P \right)}} - \tau \frac{\partial}{\partial \hat{q}_{i,m}} \sum_{m=0}^R \sum_{i=1}^N \left(\log{(\hat{q}_{i,m})} + \log{(1-\hat{q}_{i,m})}\right) \right\}\nonumber
\end{eqnarray}
Let $\mathbf{X}$ be an invertible matrix and $x$ be a scalar. Using the property
\begin{equation}
\frac{\partial \log \det{(\mathbf{X})}}{\partial x} =   tr\left\{\mathbf{X}^{-1} \frac{\partial \mathbf{X}}{\partial x}\right\}  \nonumber
\end{equation} and the definition
\begin{equation}
\mathbf{W} \triangleq \left(\sum_{m=0}^R \sum_{i=1}^N \hat{q}_{i,m} \mathbf{J}_{i,t+1}^D (m) + \mathbf{J}_{t+1}^P \right) \nonumber
\end{equation} each element of the gradient vector is obtained as
\begin{eqnarray}
\label{eq:gradient}
& (\nabla_{\hat{q}_{t+1}} \phi)_{i,m}  =  &  -tr\{\mathbf{W}^{-1}\mathbf{J}_{i,t+1}^D (m)\} - \frac{\tau}{\hat{q}_{i,m}} + \frac{\tau}{1-\hat{q}_{i,m}}
\end{eqnarray}
In order to compute $\nabla_{\hat{q}_{t+1}}^{\hat{q}_{t+1}} \phi$, for $i,i^* \in \{1,2,\ldots,N\}$ and $m,m^* \in \{0,1,\ldots,R\}$, we define
\begin{equation}
\boldsymbol{\psi}_{{(i^*,m^*)},{(i,m)}}  \triangleq \frac{\partial}{\partial \hat{q}_{i^*,m^*}} \left(\frac{\partial \mathbf{W}}{\partial \hat{q}_{i,m}}\right)  = \frac{\partial }{\partial \hat{q}_{i^*,m^*}} \left\{ -tr(\mathbf{W}^{-1}\mathbf{J}_{i,t+1}^D (m)) \right\}
\end{equation}
Using the properties
\begin{eqnarray}
&& \frac{\partial }{\partial x} tr\{\mathbf{X}\} = tr\left\{ \frac{\partial \mathbf{X} }{\partial x}  \right\}\nonumber \end{eqnarray}
and
\begin{eqnarray}
&& \frac{\partial \mathbf{X}^{-1} }{\partial x}  = -\mathbf{X}^{-1} \frac{\partial \mathbf{X}}{\partial x} \mathbf{X}^{-1} \nonumber
\end{eqnarray} we get
\begin{eqnarray}
& & \boldsymbol{\psi}_{{(i^*,m^*)},{(i,m)}} = tr\{\mathbf{W}^{-1} \mathbf{J}_{i^*,t+1}^D (m^*)\mathbf{W}^{-1} \mathbf{J}_{i,t+1}^D (m)\} \nonumber
\end{eqnarray}
Finally the Hessian matrix is obtained as
\begin{eqnarray}
\label{eq:Hessian}
&& (\nabla_{\hat{q}_{t+1}}^{\hat{q}_{t+1}} \phi) = \boldsymbol{\psi} +  \tau \: diag\left(\frac{1}{\hat{q}_{1,0}^2}+\frac{1}{(1-\hat{q}_{1,0})^2};\ldots;\frac{1}{\hat{q}_{N,R}^2}+\frac{1}{(1-\hat{q}_{N,R})^2}\right)
\end{eqnarray}

Having obtained $\nabla_{\hat{q}_{t+1}} \phi$, $\nabla_{\hat{q}_{t+1}}^{\hat{q}_{t+1}} \phi$ and ${\cal A}$, the block elimination method \cite{boyd2004convex} can be used to solve the KKT system, which is summarized in Algorithm \ref{alg:Solve_KKT}. (For the details of block elimination method see Section 10.4. in \cite{boyd2004convex}).
\begin{algorithm}                      
\caption{Solving KKT system by block elimination}          
\label{alg:Solve_KKT}                           
\begin{algorithmic}                    
\STATE (A2.1) Form $(\nabla_{\hat{q}_{t+1}}^{\hat{q}_{t+1}} \phi)^{-1}{\cal A}^T$ and $(\nabla_{\hat{q}_{t+1}}^{\hat{q}_{t+1}} \phi)^{-1}(\nabla_{\hat{q}_{t+1}} \phi)$.
\STATE (A2.2) Form ${\cal S} = - {\cal A} (\nabla_{\hat{q}_{t+1}}^{\hat{q}_{t+1}} \phi)^{-1} {\cal A}^T$
\STATE (A2.3) Determine $\omega$ by solving ${\cal S}\omega = {\cal A}(\nabla_{\hat{q}_{t+1}}^{\hat{q}_{t+1}} \phi)^{-1}(\nabla_{\hat{q}_{t+1}} \phi)$.
\STATE (A2.4) Determine $\Delta \mathbf{Z}$ by solving $(\nabla_{\hat{q}_{t+1}}^{\hat{q}_{t+1}} \phi)^{-1} \Delta \mathbf{Z} = {\cal A}^T \omega - (\nabla_{\hat{q}_{t+1}} \phi)$.
\end{algorithmic}
\end{algorithm}
In order to compute the complexity of the convex relaxation based bandwidth allocation method, we ignore the complexity for computing $(\nabla_{\hat{q}_{t+1}} \phi)$ and $(\nabla_{\hat{q}_{t+1}}^{\hat{q}_{t+1}} \phi)$. The cost of block elimination to solve (\ref{eq:Prob_BA}) is dominated by the Cholesky decomposition of $\nabla_{\hat{q}_{t+1}}^{\hat{q}_{t+1}} \phi $ which is used to find $(\nabla_{\hat{q}_{t+1}}^{\hat{q}_{t+1}} \phi)^{-1}$. Let us define ${\cal V} \triangleq N(R+1)$. Then, in order to compute the Cholesky decomposition of the matrix $(\nabla_{\hat{q}_{t+1}}^{\hat{q}_{t+1}} \phi)$, we require a total of $\frac{1}{6} ({\cal V}^3 - {\cal V})$ summations and multiplications and $\frac{1}{6}(3{\cal V}^2 - 3{\cal V})$ divisions \cite{hämmerlin1991numerical}. Thus the complexity of bandwidth allocation based on convex optimization increases with ${\cal O}(N^3(R+1)^3)$.

At each iteration of Newton's method, the solution vector $\mathbf{\hat{q}}_{t+1}$ is updated by $\mathbf{\hat{q}}_{t+1} = \mathbf{\hat{q}}_{t+1}  + s \Delta \mathbf{Z}$ where $s \in (0,1]$ is the step size obtained by the backtracking line search method \cite{boyd2004convex}. We stop the Newton iterations when the Newton decrement, $\lambda \triangleq (-\nabla_{\hat{q}_{t+1}} \phi^{T} \Delta \mathbf{Z})^{1/2}$, is less than some predefined precision. The summary of the Newton's method is presented in Algorithm \ref{alg:Newton_bw}.

\begin{algorithm}                      
\caption{Newton's method for the bandwidth allocation problem}          
\label{alg:Newton_bw}                           
\begin{algorithmic}                    
\STATE Find a feasible starting point $\mathbf{\hat{q}}_{t+1}$ from (\ref{eq:linprog}) and set precision $\epsilon>0$
\STATE Repeat
\STATE (A3.1) Compute the Newton Step $\Delta \mathbf{Z}$ from (\ref{eq:Newton_KKT}) and Newton decrement $\lambda = (-\nabla_{\hat{q}_{t+1}} \phi^{T} \Delta \mathbf{Z})^{1/2}$
\STATE (A3.2) Choose Step size $s$ by backtracking line search.
\STATE (A3.3) Update $\mathbf{\hat{q}}_{t+1} = \mathbf{\hat{q}}_{t+1}  + s \Delta \mathbf{Z}$
\STATE (A3.4) Quit if $\lambda^2 / 2 \leq \epsilon$, then $\mathbf{\hat{q}}^{*}_{t+1} = \mathbf{\hat{q}}_{t+1}$, else go to Step (A3.1).
\end{algorithmic}
\end{algorithm}

\subsubsection*{The Probabilistic Transmission Scheme}

From the optimal solution of the relaxed problem $\mathbf{\hat{q}}^{*}_{t+1}$, the bandwidth distribution for the next time step needs to be determined. For the sensor selection problem, the authors in \cite{sp09:boyd} employed a simple scheme in which $k$ sensors are selected out of $N$ sensors by first sorting $\mathbf{\hat{q}}^{*}_{t+1}$ in descending order and then setting $k$ largest elements of $\mathbf{\hat{q}}^{*}_{t+1}$ to one. For the bandwidth distribution problem, a similar solution is to sort the probabilities $\hat{q}_{i,m}$ in descending order and then to assign 1 starting from the largest probability until the bandwidth constraint is satisfied. However, in this paper, we consider a randomized scheme similar to the ones used in \cite{sp09:boyd} and \cite{masazade:ciss10}. Since the elements of $\mathbf{\hat{q}}^{*}_{t+1}$ are within the range $[0,1]$, we can consider each $\hat{q}_{i,m}$ as the transmission probability of sensor $i$, transmitting information in $m$ bits. Instead of putting a strict bandwidth constraint, i.e. $\sum_{i=1}^N R_{i,t+1}  = R$, the probabilistic transmission puts a weak constraint on the bandwidth availability and ensures that the sensors on the average transmit $R$ bits to the fusion center. We present a numerical example on the probabilistic transmission scheme in Section \ref{section:Sim_Results}.

\subsection{Approximate Dynamic Programming based Bandwidth Distribution}
\label{sec3:dp_intro}

In this section, we present the bandwidth allocation algorithm based on A-DP which will be shown to provide near optimal solution but require much less computation time than the convex relaxation approach. Note that the Fisher information matrix can be expressed as the summation of each sensor's individual Fisher information matrices as defined in (\ref{eq:FIM_BA}).  In this section, we formulate an approximate DP recursion in tracking applications where we can maximize the Fisher information by maximizing its determinant subject to the bandwidth constraint.

\begin{figure}[hbt]
\centerline{ \begin{tabular}{c}
\includegraphics[width=.5\textwidth,height=!]{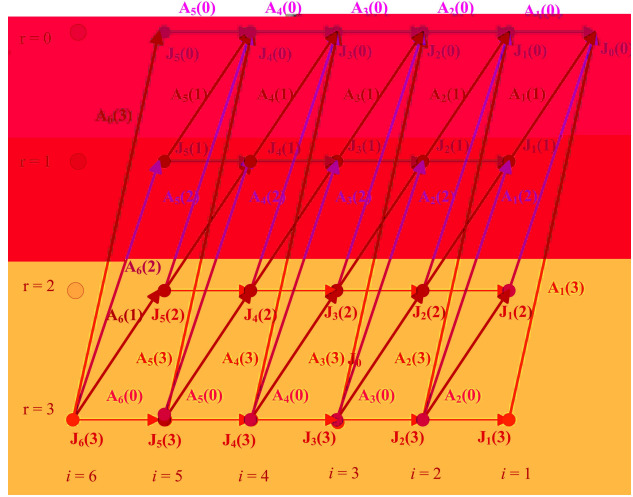}  \\
\end{tabular}}
      \caption{Trellis of the DP for tracking time step $t$. $(N=6, \: R=3)$.}
  \label{fig:DP_Trellis}
\end{figure}

Typically DP involves progression along time. But in our problem formulation the DP progresses across sensors and is executed at each time step of tracking to determine the bandwidth allocations of the next time step. Since A-DP is performed at each time step, for simplicity, the time index $t+1$ for Fisher
information matrix is dropped. Instead an index for the stages in DP is adopted. Let $\mathbf{J}_N = \mathbf{J}_{t+1}$ and $\mathbf{A}_{i}(R_i) = \mathbf{J}_{t+1}^D (R_i)$ be the reward in terms of Fisher information when sensor $i$ quantizes its measurement in $R_i$ bits ($R_i \in \{0,1,\ldots, R\}$). While constructing the DP trellis, the bandwidth distribution problem is first divided into $N+1$ stages which correspond to $N$ sensors and a termination stage. We define the state of a stage as the remaining bandwidth for the usage of sensor $i$.  So each stage has $R+1$ states associated with it. The bandwidth allocation chosen at any sensor (stage) determines the feasible states at the next sensor. An example DP trellis is shown in Fig. \ref{fig:DP_Trellis} with $N = 6$ and $R=3$ which implies a total of 7 stages and 4 states in the DP trellis.  As an example, sensor $1$ is at state $r=1$ means 2 bits have already been used by the other $N-1$ sensors and 1 bit is available for sensor $1$. Then, sensor 1 can only take the action $\mathbf{A}_1(1)$ and the DP goes to the termination stage (stage $0$) which has only the 0 bit available state.

For such a DP trellis, we have,
\begin{eqnarray}
\label{eq:DP_recurs1}
& \mathbf{J}_N & = \mathbf{A}_{N}(R_N) + \left\{\mathbf{A}_{N-1}(R_{N-1}) + \ldots + \mathbf{A}_{1}(R_1) + \mathbf{J}_{0} \right\} \nonumber \\
&&  = \mathbf{A}_{N}(R_N) + \mathbf{J}_{N-1} \nonumber \\
&& : \nonumber \\
& \mathbf{J}_1 & = \mathbf{A}_{1}(R_1) + \mathbf{J}_{0}
\end{eqnarray} where $\mathbf{J}_{0} = \boldsymbol{\Sigma}_{t+1}^{-1}$ and $\sum_{i=1}^N R_{i} = R$. According to the matrix determinant lemma \cite{petersen2008matrix}, $$\det(\mathbf{X}+\mathbf{A}) = \det(\mathbf{X}+\mathbf{A}\mathbf{I}) = \det(\mathbf{X})\det(\mathbf{I}+\mathbf{X}^{-1}\mathbf{A})$$
With $\mathbf{X} = \mathbf{J}_{i-1}$, $\mathbf{A} = \mathbf{A}_i(R_i)$, and $\mathbf{I}$ being the identity matrix, we have \begin{eqnarray}
&& \log \Big\{ \det(\mathbf{J}_N) \Big\} =  \log\Big\{\det(\mathbf{J}_{N-1})\Big\} + \log \Big\{\det\left[\mathbf{I} + \mathbf{J}_{N-1}^{-1}\mathbf{A}_{N}(R_{N})\right] \Big\}\nonumber \\
&& : \nonumber \\
&& \log \Big\{ \det(\mathbf{J}_{1}) \Big\} =  \log \Big\{\det(\mathbf{J}_{0}) \Big\} + \log \Big\{\det(\mathbf{I} + \mathbf{J}_{0}^{-1}\mathbf{A}_{1}(R_{1})) \Big\}\nonumber
\end{eqnarray}

We can maximize $\det(\mathbf{J}_N)$, by maximizing $\log\Big\{\det(\mathbf{J}_N)\Big\}$. The DP recursion at each stage is formulated as follows: the trellis starts from $\mathbf{J}_0(0)$ and for the first stage ($i=1$) and $\textrm{for all} \: r \in \{0,1,\ldots,R\}$,
\begin{eqnarray}
\label{eq:DP_recursuions0}
&& \log \Big[\det[\mathbf{J}_1(r)]\Big] = \log\Big[\det[\mathbf{I} + \mathbf{J}_{0}^{-1}(0)\mathbf{A}_1(r)]\Big]  + \log\Big[\det[\mathbf{J}_{0}(0)]\Big]
\end{eqnarray}
Then for all the intermediate stages $i \in \{2,\ldots,N-1\} $ and $\textrm{for all}\: r \in \{0,1,\ldots,R\}$,
\begin{eqnarray}
\label{eq:DP_recursuions1}
&& \log \Big[\det[\mathbf{J}_i(r)]\Big] =  \\
&& \max_{k=0,1,\ldots,r} \: \bigg\{ \log\Big[\det[\mathbf{I} + \mathbf{J}_{i-1}^{-1}(r-k)\mathbf{A}_i(k)]\Big]  +
 \log\Big[\det[\mathbf{J}_{i-1}(r-k)]\Big]\bigg\} \nonumber
\end{eqnarray}
Finally for the last stage $i=N$,
\begin{eqnarray}
\label{eq:DP_recursuions2}
&& \log \Big[\det[\mathbf{J}_N(R)]\Big] =  \\
&& \max_{k=0,1,\ldots,R} \: \bigg\{ \log\Big[\det[\mathbf{I} + \mathbf{J}_{N-1}^{-1}(R-k)\mathbf{A}_i(k)]\Big]  +  \log\Big[\det[\mathbf{J}_{N-1}(R-k)]\Big]\bigg\} \nonumber
\end{eqnarray}

In (\ref{eq:DP_recursuions0}), (\ref{eq:DP_recursuions1}), and (\ref{eq:DP_recursuions2}), the reward of
sensor $i$'s transmission in $R_i$ bits depends not only on $\mathbf{A}_i(R_i)$ but also on the FIM of the previous stage $\mathbf{J}_{i-1}^{-1}$. So at each stage $i$, the FIM, $\mathbf{J}_i(r)$,  which has the maximum determinant should be stored in a memory for its use at the next recursion. Note that the proposed A-DP may not yield the maximum matrix determinant at the final stage. The suboptimality of the A-DP recursions is discussed later in this section.

We analyze the computational complexity of A-DP in terms of number of matrix summations. Note that the number of element-wise summation is a scaled version of number of matrix summations. The first stage needs $R$ matrix summations to compute the FIM at all states. For all the intermediate stages, at each state $r$, $(r \in \{1,\ldots,R\})$, $r$ different matrix summations are required to find the FIM with the maximum determinant. Finally at stage $N$,  A-DP again needs over $R$ matrix summations in order to maximize the determinant of $\det{\mathbf{J}_N}$. So the A-DP totally searches over,
\begin{eqnarray}
&& R + (N-2)\left[\sum_{r=1}^R r \right] + R = 2R + (N-2)\frac{R(R+1)}{2} \nonumber
\end{eqnarray} matrix summations which is linear in $N$ and quadratic in $R$.

\subsubsection*{Suboptimality of the A-DP recursions}
For a given state of a stage, we choose the path with the maximum determinant of the FIM and dismiss all the other paths
arriving at this state. The proposed DP recursions would yield the optimal solution to maximize the determinant
of the FIM, if the following property were satisfied,
\begin{eqnarray}
&& \mathrm{if} \quad \det\{\mathbf{J}^{'}\} \geq \det\{\mathbf{J}^{''}\} \\
&& \mathrm{then} \quad \det\{\mathbf{A} + \mathbf{J}^{'}\} \geq \det\{\mathbf{A} + \mathbf{J}^{''}\} \nonumber
\end{eqnarray} for some positive semidefinite matrices $\mathbf{J}^{'}$, $\mathbf{J}^{''}$ and $\mathbf{A}$.
Unfortunately, the above property is not necessarily true. Consider the simple example, $\mathbf{J}^{'} = \left(
                                                                               \begin{array}{cc}
                                                                                 1 & 0 \\
                                                                                 0 & 1 \\
                                                                               \end{array}
                                                                             \right)$ and $\mathbf{J}^{''} = \left(
                                                                               \begin{array}{cc}
                                                                                 1 & -0.1 \\
                                                                                 -0.1 & 1 \\
                                                                               \end{array}
                                                                             \right)$ where $\det\{\mathbf{J}^{'}\} > \det\{\mathbf{J}^{''}\}$. Let $\mathbf{A} = \left(
                                                                                                         \begin{array}{cc}
                                                                                                           1 & 0.1 \\
                                                                                                           0.1 & 1 \\
                                                                                                         \end{array}
                                                                                                       \right)$. Then
                                                                                                       $\det\{\mathbf{A} + \mathbf{J}^{'}\}< \det\{\mathbf{A} + \mathbf{J}^{''}\}$.

At each stage of the DP, we only store the FIM with the maximum determinant. Therefore, the final solution obtained
by the DP recursions becomes suboptimal since not all the feasible solutions are enumerated.

\subsection{Existing Suboptimal Bandwidth Distribution Methods}

In this section, we review some existing suboptimal methods that are suitable for solving the bandwidth
allocation problem in target tracking applications.

\subsubsection{GBFOS Algorithm}

This algorithm has been first proposed in \cite{icassp10:onur} for dynamic bandwidth allocation in target tracking. The GBFOS algorithm starts by assigning the maximum number of bits, $R$ to each sensor in the network and then reduces the number of bits one bit at a time until the sum rate constraint is satisfied. The GBFOS algorithm can be stated as in Algorithm \ref{alg3}. Note that in order to simplify the notation, we drop the time index $t+1$ in the algorithm. As shown in Step (A4.1), at each iteration, the GBFOS algorithm searches the $N$ sensors and reduces the bits of the sensor by one which ensures the minimum reduction of the determinant of the FIM. An efficient implementation of the GBFOS algorithm and its complexity analysis can be given as follows:

Let us define $\mathbf{J}^{i}$ as the FIM after the $i^{th}$ iteration, and $\mathbf{A}_k(R_k) \triangleq \mathbf{J}_{t+1}^D (R_k) $ as sensor $k$'s contribution to the FIM using $R_k$ bits. In the beginning, we need to calculate $\mathbf{J}^{0} = \mathbf{J}_{t+1}^P+ \mathbf{A}_1(R)+\mathbf{A}_2(R)+....+\mathbf{A}_N(R)$. As a result, totally $N$ matrix summations are needed. At the $i$-th iteration, where $i=1, \ldots, (N-1)R$, there are at most $N$ different ways to reduce 1 bit. Assuming one particular solution is to reduce one bit at the $k$-th sensor, ($k \in \{1,2,\ldots,N\}$), and $\mathbf{J}^{i}(k)=\mathbf{J}^{i-1}+\mathbf{A}_k(R_k-1)-\mathbf{A}_k(R_k)$, which requires two matrix summations. Hence at each iteration, at most $2N$ matrix summations are required. At the end of the $i^{th}$ iteration, we store $\mathbf{J}^{i}(k)$ with the maximum determinant. In summary, we need at most $N+2N(N-1)R$ matrix summations, which is quadratic in $N$ and linear in $R$. Note that this is an upper bound on complexity.

\begin{algorithm}                      
\caption{GBFOS - Bandwidth Distribution Algorithm}          
\label{alg3}                           
\begin{algorithmic}                    
\STATE Set $\mathbf{R}_0 = [R_1 = R, \ldots, R_N = R]$ and $\mathbf{J}^0 = \mathbf{J}_{t+t}^P+ \mathbf{A}_1(R)+\mathbf{A}_2(R)+....+\mathbf{A}_N(R)$.
\STATE \textbf{FOR} $i = 1:(N-1)R$
\qquad \STATE (A4.1) \textbf{FOR} $k = 1:N$ \\
\qquad\qquad \textbf{IF} {$R_k > 0$} \\
\qquad\qquad\qquad    Reduce one bit from sensor $k$ and compute $\det(\mathbf{J}^{i}(k))$ where \\
\qquad\qquad\qquad $\mathbf{J}^{i}(k)=\mathbf{J}^{i-1}+\mathbf{A}_k(R_k-1)-\mathbf{A}_k(R_k)$. \\
\qquad\qquad    \textbf{ENDIF}\\
\qquad    \textbf{ENDFOR}
\STATE (A4.2) $\forall \: k$ with $R_k > 0$, find the sensor $p^*$ for which $\det(\mathbf{J}^{i}(k))$ is the maximum: \\
\qquad ${p^*} = \displaystyle \arg \max_{k \: \mathrm{where} \: R_k > 0} \: \det(\mathbf{J}^{i}(k))$.
\STATE (A4.3) Decrement $R_{p^*} = R_{p^*}-1$, update $\mathbf{R}_{i} = [R_1, \ldots, R_{p^*},\ldots,R_N]$ and set $\mathbf{J}^i=\mathbf{J}^i(p^*)$.
\STATE \textbf{ENDFOR}
\end{algorithmic}
\end{algorithm}

\subsubsection{Greedy Algorithm}

Basically, greedy search is the reverse of the GBFOS method which makes the algorithm much faster. The greedy algorithm can be stated as in Algorithm \ref{alg4}. The greedy algorithm starts by assigning $0$ bits to each sensor in the network and then increases the number of bits one bit at a time until the sum rate constraint is satisfied in $R$ iterations. At each iteration, greedy algorithm searches the $N$ sensors and a single bit is added to the sensor which maximizes the determinant of the resulting FIM.

The implementation of greedy search and its complexity can be stated as follows: At the first iteration, there are $N$ different ways to add 1 bit. For the $k$-th way of adding 1 bit, $\mathbf{J}^{1}(k)=\mathbf{J}^P_{t+1}+\mathbf{A}_k(1)$. Then we set $\mathbf{J}^1 = \max_{k} \det ( \mathbf{J}^{1}(k) )$. Hence, totally $N$ matrix summations are required at the first iteration. At the $i$-th iteration, for $i=2,\ldots,R$, there are still $N$ different ways to add  1 bit. For the $k$-th way of adding 1 bit, $\mathbf{J}^i(k)=\mathbf{J}^{i-1}+\mathbf{A}_k(R_k+1)-\mathbf{A}_k(R_k)$.  Note that $\mathbf{A}_k(R_k)$ could be a zero matrix since $R_k$ could be zero. Therefore, for each iteration, at most a total of $2N$ matrix summations are required. In summary, we need at most $N+(R-1)2N=N(2R-1)$ matrix summations which is an upper bound for the complexity of greedy search.

\begin{algorithm}                      
\caption{Greedy Bandwidth Distribution Algorithm}          
\label{alg4}                           
\begin{algorithmic}                    
\STATE Set $\mathbf{R}_0 = [R_1 = 0, \ldots, R_N = 0]$ and $\mathbf{J}^0 = \mathbf{J}_{t+1}^P$.
\STATE \textbf{FOR} $i = 1:R$
\qquad \STATE (A4.1) \textbf{FOR} $k = 1:N$ \\
\qquad\qquad Add one bit to sensor $k$ and compute $\det(\mathbf{J}^{i}(k))$ where \\
\qquad\qquad $\mathbf{J}^{i}(k)=\mathbf{J}^{i-1}+\mathbf{A}_k(R_k+1)-\mathbf{A}_k(R_k)$. \\
\qquad    \textbf{ENDFOR}
\STATE (A4.2) Find the sensor $p^*$ for which $\det(\mathbf{J}^{i}(k))$ is the maximum: \\
\qquad ${p^*} = \displaystyle \arg \max_{k} \: \det(\mathbf{J}^{i}(k))$.
\STATE (A4.3) Increment $R_{p^*} = R_{p^*}+1$, update $\mathbf{R}_{i} = [R_1, \ldots, R_{p^*},\ldots,R_N]$ and set $\mathbf{J}^i=\mathbf{J}^i(p^*)$.
\STATE \textbf{ENDFOR}
\end{algorithmic}
\end{algorithm}

\section{Simulation Results}
\label{section:Sim_Results}

In this section, we illustrate the performance of different dynamic bandwidth distribution methods with numerical examples. For the convex relaxation problem, we solve the linear programming problem in (\ref{eq:linprog}) which is formulated to find the feasible initial point by using the ``linprog'' routine in MATLAB . The Newton method parameters $\epsilon$ and $\tau$ are selected according to \cite{cnvx_code1}. Simulation results show that, for the convex relaxation scheme, the optimal solution is reached in around ten iterations.

\begin{figure}[hbt]
\centerline{ \begin{tabular}{c}
\includegraphics[width=.5\textwidth,height=!]{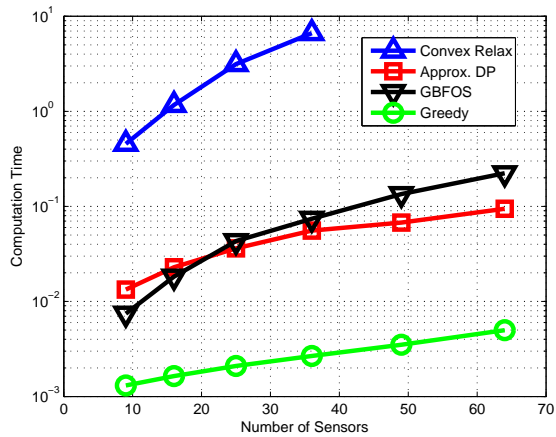} \\
\end{tabular}}
      \caption{Computation time in seconds for convex relaxation, A-DP, GBFOS, and greedy search $(R=5)$.}
  \label{fig:Comp_Time}
\end{figure}

We evaluate the computation time of each bandwidth allocation approach by using the ``etime'' function of MATLAB  averaged over 100 trials. In Fig. \ref{fig:Comp_Time}, the mean computation times of the considered suboptimal bandwidth allocation schemes are compared. Since the number of summations for A-DP increases linearly with $N$, for large number of sensors,  the computation time of A-DP is less than the computation time of GBFOS and convex relaxation
where the number of summations increase quadratically and cubically respectively.

\begin{figure*}[hbt]
\centerline{ \begin{tabular}{cc}
\includegraphics[width=.5\textwidth,height=!]{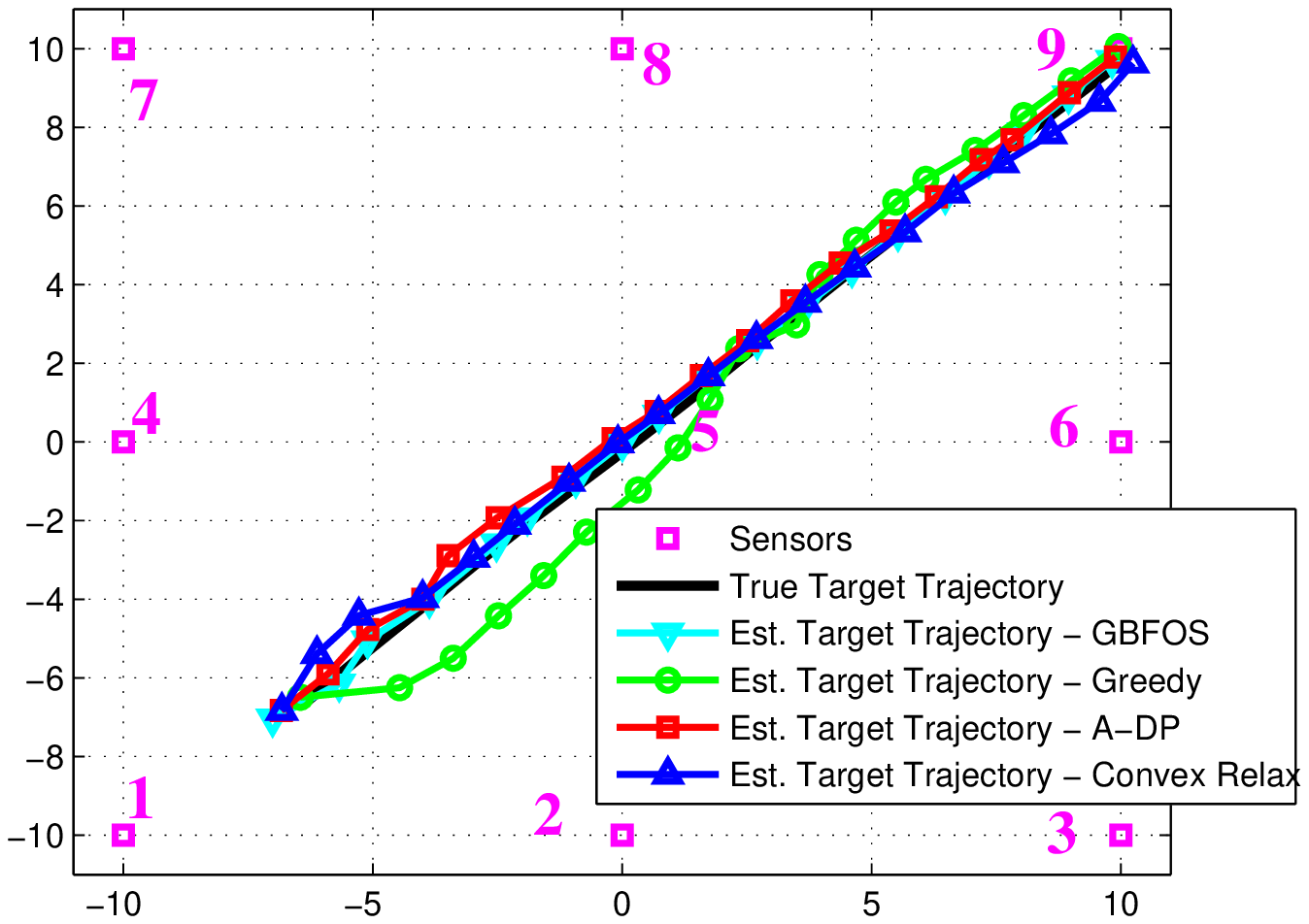} & \includegraphics[width=.5\textwidth,height=!]{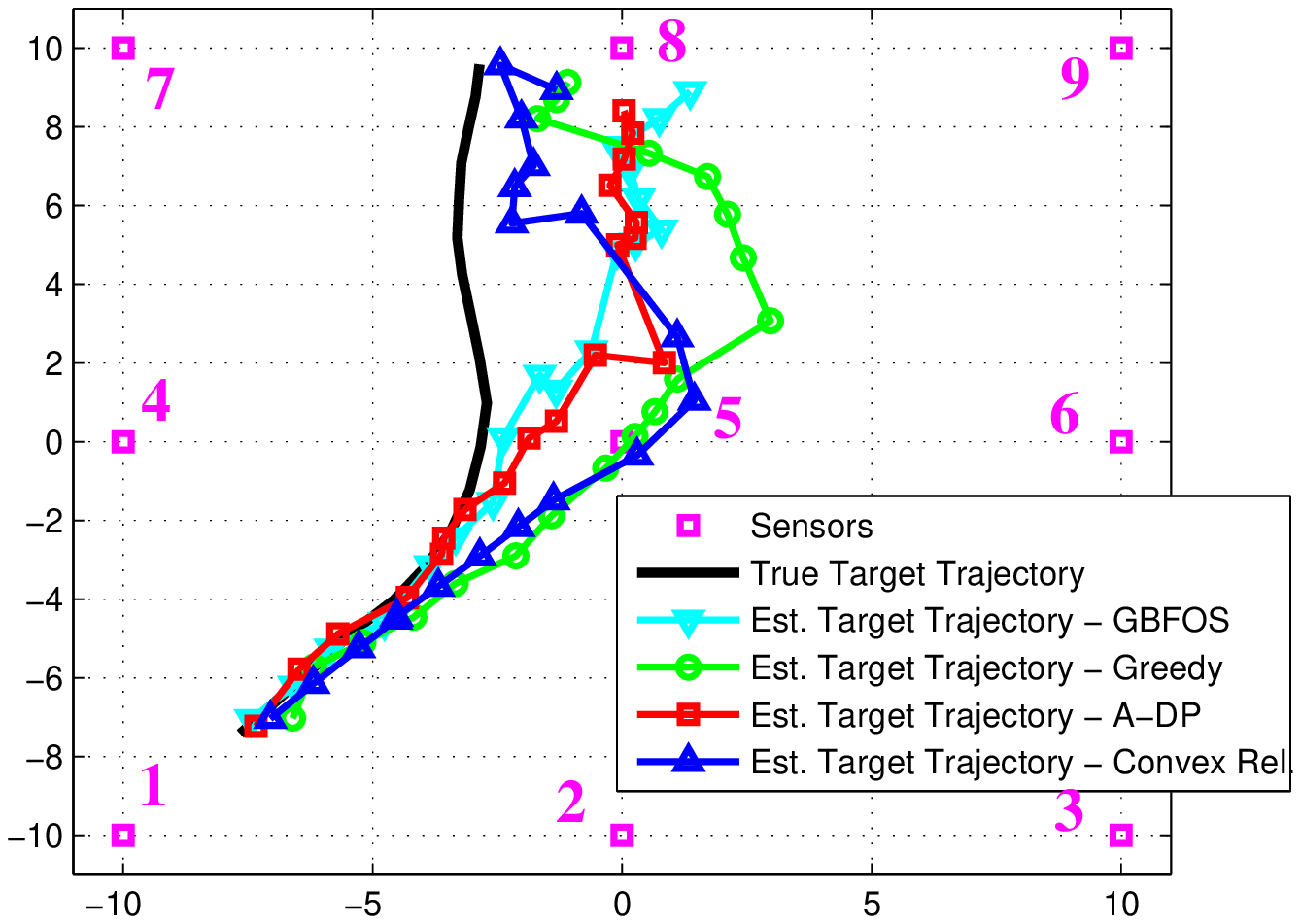} \\
(a) & (b) \\
\end{tabular}}
      \caption{A WSN with $N=9$ sensors tracking sample targets (a) $\rho = 2.5\times10^{-3}$ (b) $\rho = 0.1$.}
  \label{fig:Res_SimIntro}
\end{figure*}

We next compare the MSE performances of the bandwidth distribution schemes based on optimal exhaustive search, A-DP, convex optimization, GBFOS and greedy search. In addition, we analyze the MSE performance of nearest neighbor bandwidth allocation, where all the bandwidth is assigned to the sensor which is nearest to the predicted target location. In our simulations, we assume that $N$ sensors are grid deployed in a $b^2 = 20\:m\; \times 20 \:m$ surveillance area as shown in Fig. \ref{fig:Res_SimIntro}-(a) and (b). We select $P_0 = 10^3$ and sensor observation noise $\sigma^2 = 1$. The probability density function of the target's initial state, $p(\mathbf{x}_0)$, is assumed to be Gaussian with mean $\mu_0 = [-8\;-8\;  2\;2]$ and covariance $\Sigma_0 = diag[\sigma_{\theta}^2\; \sigma_{\theta}^2\; 0.01\; 0.01]$ where we select $3\sigma_{\theta} = 2$ so the initial point of the target remains in the ROI with very high probability. The target motion follows a white noise acceleration model and we consider two process noise parameters $\rho = 2.5 \times 10^{-3}$ and $\rho = 0.1$. Measurements are assumed to be taken at regular intervals of ${\cal D} = 0.5$ seconds and the observation length is $10$ seconds. Namely, we perform target tracking over $T_S = 20$ time steps for each Monte-Carlo trial. The number of particles used in the particle filter is $N_s = 5000$. We assume $R=5$ bits of bandwidth is available at each time step for data transmission.  The MSE at each time step is averaged over $T_{trials} = 500$ trials as,
\begin{eqnarray}
&& \textrm{MSE}(t) = \frac{1}{T_{trials}} \sum_{v=1}^{T_{trials}}  \left[(\mathbf{x}_{t}^v(1)-\mathbf{\hat{x}}_{t}^v(1))^2+(\mathbf{x}_{t}^v(2)-\mathbf{\hat{x}}_{t}^v(2))^2 \right]
\end{eqnarray} where in the $v^{th}$ trial $\mathbf{x}_{t}^v$ and $\mathbf{\hat{x}}_{t}^v$ are the actual and
estimated target states at time $t$ respectively.

In Fig. \ref{fig:Res_SimIntro}-(a) and (b), a WSN is illustrated where $N=9$ sensors track a target under the process noise parameters $\rho = 2.5 \times 10^{-3}$ and $\rho = 0.1$ respectively. For $\rho =  2.5 \times 10^{-3}$, the process noise is relatively small and the target trajectory is almost deterministic. For $\rho = 0.1$, the target trajectory has relatively large uncertainty. For the first time step of tracking, Table \ref{table:optimal_rates} presents each sensor's transmission probability for each quantization rate for the convex optimization based bandwidth allocation scheme with $R=5$. Note that at $t=1$, the target is relatively close to sensor 1 located at $(-10\:m.,-10\:m.)$. Then it is very likely that sensor $1$ transmits its measurement using $m=5$ bits because of the probability, $\hat{q}_{1,5} \approx 0.84$. Rest of the sensors tend to remain silent since their transmission probabilities using $0$ bits are almost 1. As seen in Table \ref{table:Sta_ConvX}, the probabilistic transmission introduces a weak constraint on the bandwidth and on the average sensors transmit $R$ bits to the fusion center.

\begin{table}[hbt]
\caption{Transmission probabilities of each quantization rate for $N=9$ and $R=5$ at $t=1$ for the example illustrated in Fig. \ref{fig:Res_SimIntro}-(a).}
    \label{table:optimal_rates}
\begin{center}
    \begin{tabular}{ | c | c | c | c | c | c | c |}
    \hline
        & $m=0$  & $m=1$ & $m=2$ & $m=3$ & $m=4$  & $m=5$ \\ \hline
$i=1$ &  0.0021  &  0.0010 &   0.0011  &  0.0037  &  0.1482  &  $\mathbf{0.8440}$ \\ \hline
$i=2$ &  $\mathbf{0.9877}$  &  0.0082 &   0.0003  &  0.0017  &  0.0012  &  0.0009 \\ \hline
$i=3$ &  $\mathbf{0.9895}$  &  0.0053 &   0.0021  &  0.0014  &  0.0010  &  0.0008 \\ \hline
$i=4$ &  $\mathbf{0.9906}$  &  0.0031 &   0.0031  &  0.0017  &  0.0011  &  0.0003 \\ \hline
$i=5$ &  $\mathbf{0.9888}$  &  0.0057 &   0.0023  &  0.0014  &  0.0010  &  0.0008 \\ \hline
$i=6$ &  $\mathbf{0.9895}$  &  0.0053 &   0.0021  &  0.0014  &  0.0010  &  0.0008 \\ \hline
$i=7$ &  $\mathbf{0.9895}$  &  0.0053 &   0.0021  &  0.0014  &  0.0010  &  0.0008 \\ \hline
$i=8$ &  $\mathbf{0.9895}$  &  0.0053 &   0.0021  &  0.0014  &  0.0010  &  0.0008 \\ \hline
$i=9$ &  $\mathbf{0.9895}$  &  0.0053 &   0.0021  &  0.0014  &  0.0010  &  0.0008 \\ \hline
\end{tabular}
\end{center}
\end{table}

\begin{table}[hbt]
\caption{Mean and Standard deviation of the total number of transmitted bits using convex relaxation based bandwidth allocation method, $R=5$.}
    \label{table:Sta_ConvX}
\begin{center}
    \begin{tabular}{ | l | c | c |}
    \hline
        & Mean  & Standard Deviation \\ \hline
$N=9$, $\rho = 2.5 \times 10^{-3}$ &  5.0071  & 1.9781 \\ \hline
$N=25$, $\rho = 2.5 \times 10^{-3}$& 5.0291  & 1.6264 \\ \hline
$N=9$, $\rho = 0.1$ & 5.0129 &  1.3439 \\ \hline
$N=25$, $\rho = 0.1$ & 5.0093 &  1.2119 \\ \hline
\end{tabular}
\end{center}
\end{table}


\begin{figure*}[hbt]
\centerline{ \begin{tabular}{cc}
\includegraphics[width=.5\textwidth,height=!]{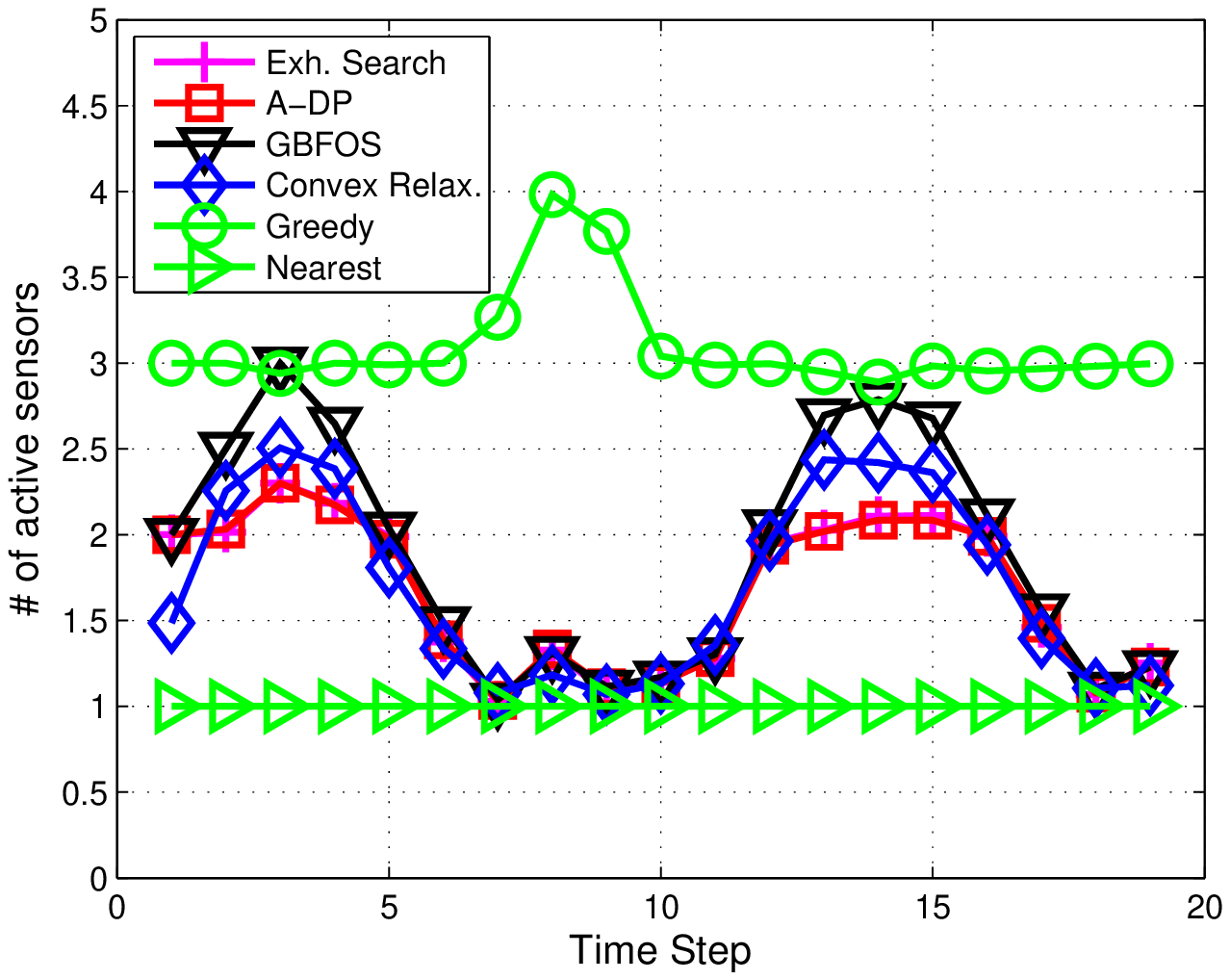} & \includegraphics[width=.5\textwidth,height=!]{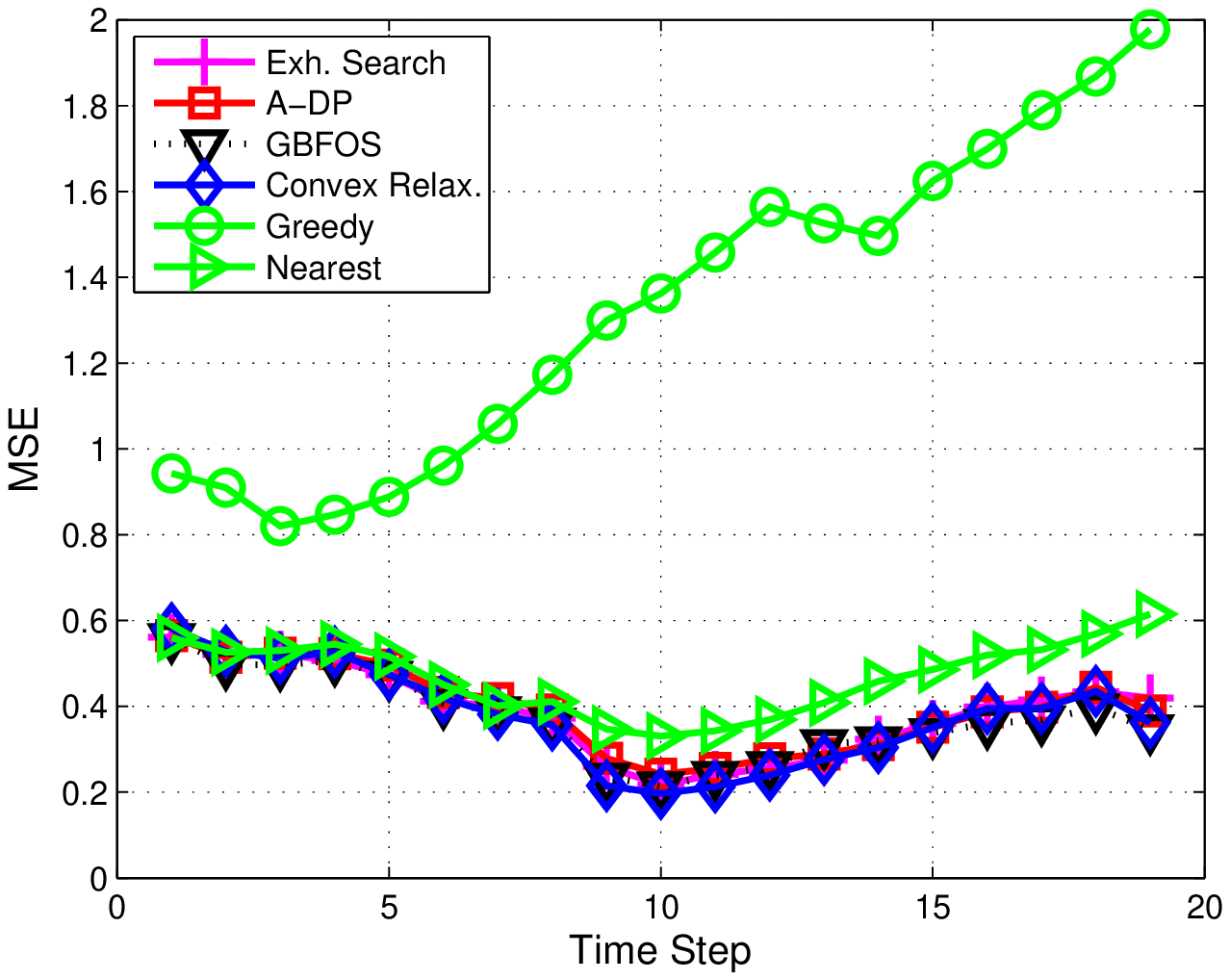} \\
(a) & (b)\\
\includegraphics[width=.5\textwidth,height=!]{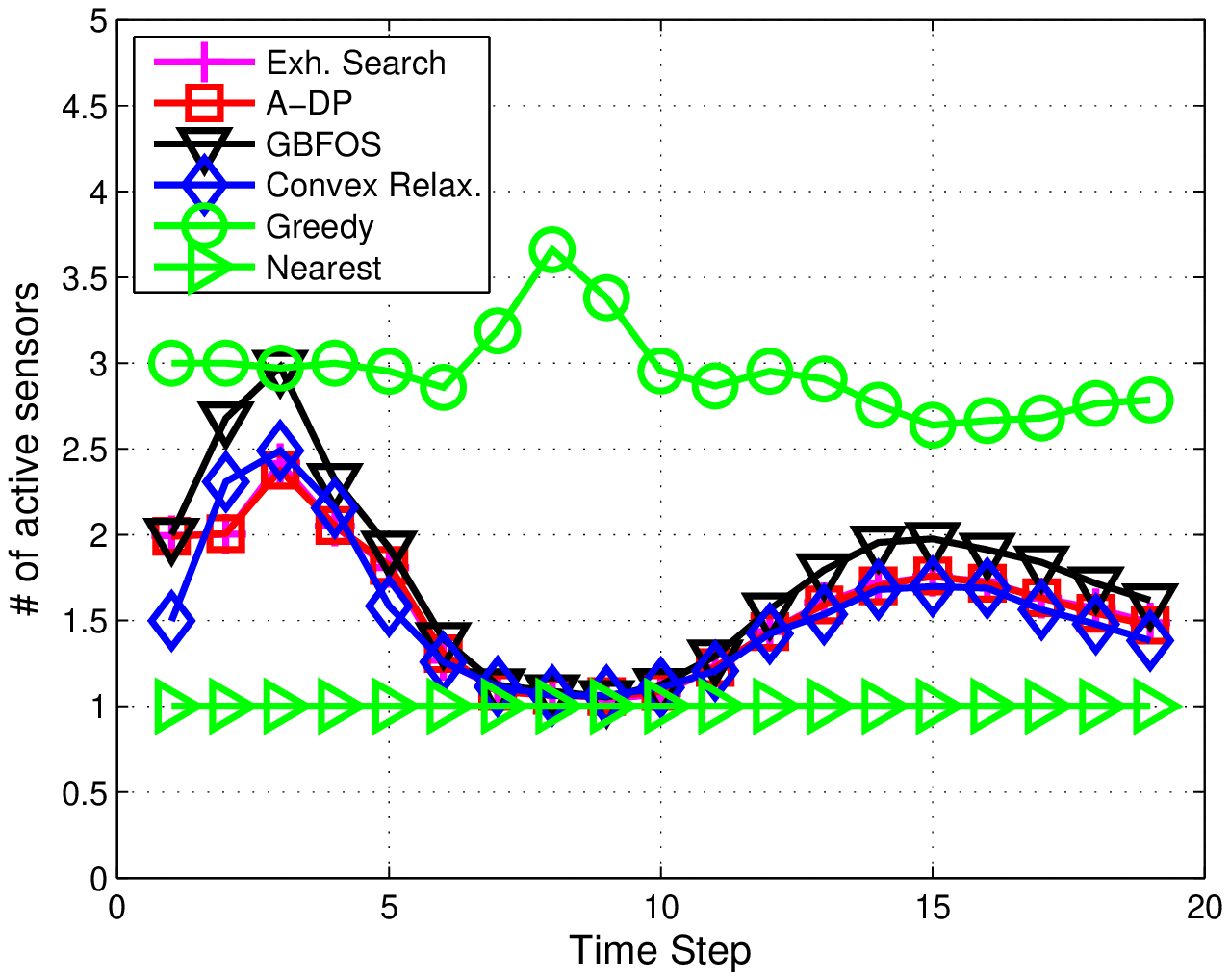} & \includegraphics[width=.5\textwidth,height=!]{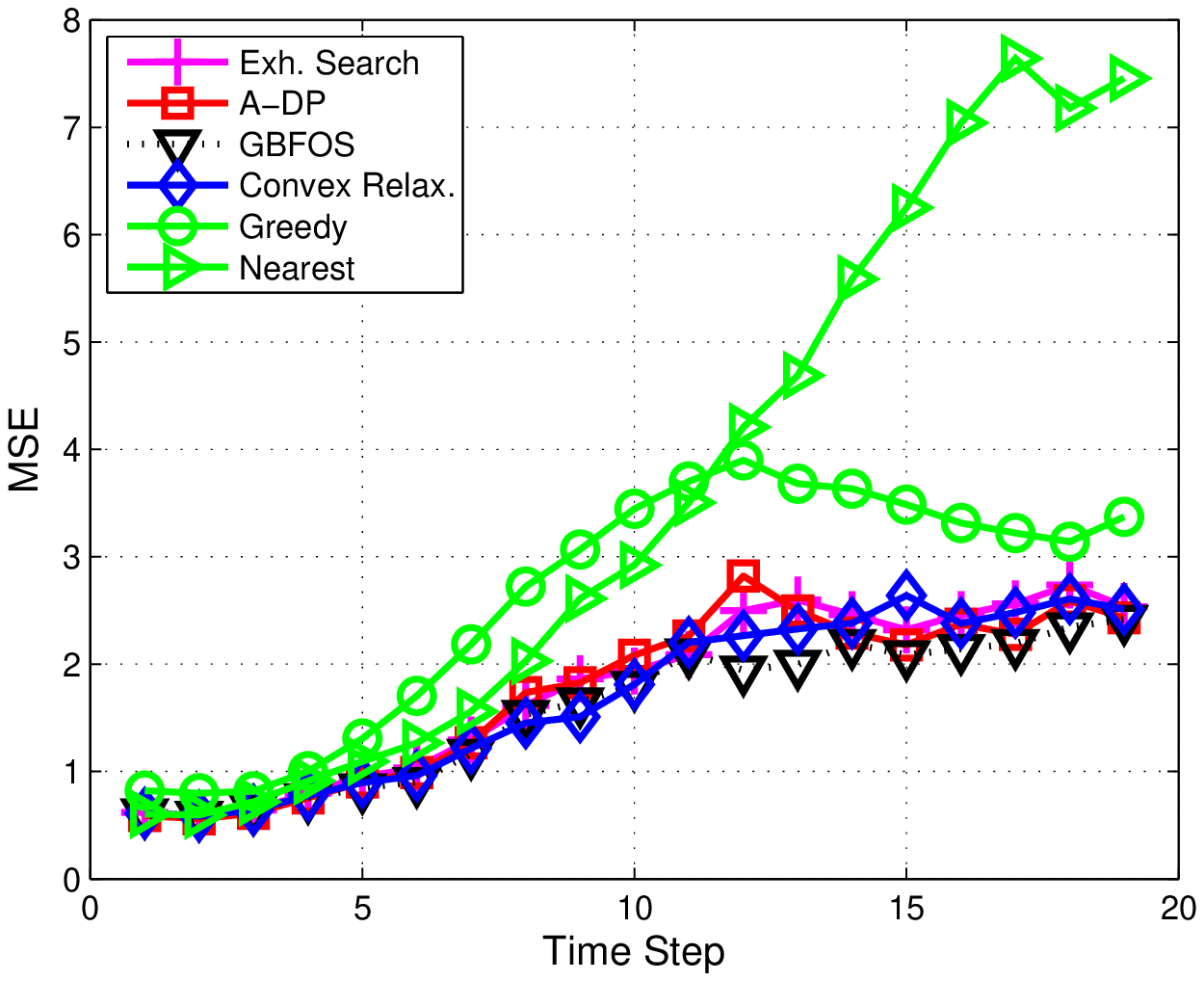}\\
(c) & (d)\\
\end{tabular}}
      \caption{ $N=9$, $R=5$, $T_{trial} = 500$  (a) Average number of active sensors,  $\rho = 2.5\times 10^{-3}$ , (b) MSE at each time step, $\rho = 2.5\times10^{-3}$, (c) Average number of active sensors,  $\rho = 0.1$, (d) MSE at each time step, $\rho =0.1$.}
  \label{fig:Act_Sens_MSE_N9}
\end{figure*}

For $N=9$ sensors, Figs. \ref{fig:Act_Sens_MSE_N9}-(a) and \ref{fig:Act_Sens_MSE_N9}-(c) show the average number of sensors activated and Figs. \ref{fig:Act_Sens_MSE_N9}-(b) and \ref{fig:Act_Sens_MSE_N9}-(d) show the MSE at each time step of tracking averaged over 500 Monte-Carlo trials. Simulation results show that under $\rho = 2.5 \times 10^{-3}$, convex relaxation, A-DP and GBFOS yield similar tracking performance to that of exhaustive search in terms of MSE.  For $\rho = 2.5 \times 10^{-3}$, between the time steps 8 and 10, the target is relatively close to sensor $5$ located at $(0,0)$. Hence, using exhaustive search, A-DP, convex relaxation, and GBFOS based bandwidth allocation schemes, almost all the bandwidth is allocated to sensor $5$. When the target is not relatively close to any of the sensors, as in time steps 2-6 and 12-17, the fusion center has relatively large uncertainty about the target location, so multiple sensors are activated with relatively coarse information which increases the estimation error as shown in Fig. \ref{fig:Act_Sens_MSE_N9}-(b). After time step 17, the target approaches sensor 9 and by using exhaustive search, convex relaxation, A-DP and GBFOS, all the bandwidth is assigned to sensor 9 and then estimation error reduces again. The greedy bandwidth allocation scheme tends to activate more sensors all the time with relatively coarse information as compared to the other bandwidth allocation algorithms. With small process noise parameter $(\rho = 2.5 \times 10^{-3})$, nearest neighbor based bandwidth allocation becomes more accurate than greedy search since the target trajectory is highly deterministic and there is a small uncertainty on the predicted target location. However, the tracking performance of the nearest neighbor approach is still not as good as those for exhaustive search, convex relaxation, GBFOS, and A-DP. For $\rho = 0.1$, the uncertainty on target trajectory is relatively large and we observe a worse tracking performance as compared to the $\rho = 2.5 \times 10^{-3}$ case. On the other hand, still A-DP, convex optimization, and GBFOS perform equally well as exhaustive search in terms of MSE and outperform greedy search. For $\rho = 0.1$, nearest neighbor based bandwidth allocation introduces much larger estimation errors which are sometimes even greater than those obtained by the greedy search based dynamic bandwidth allocation scheme.

\begin{figure*}[hbt]
\centerline{ \begin{tabular}{cc}
\includegraphics[width=.5\textwidth,height=!]{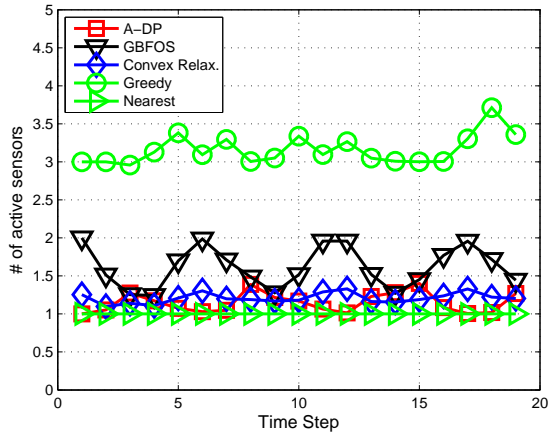} & \includegraphics[width=.5\textwidth,height=!]{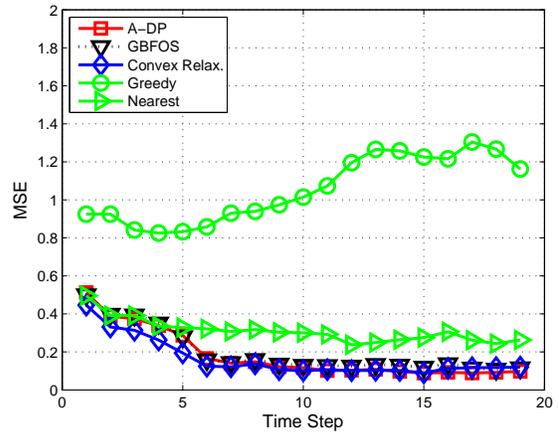} \\
(a) & (b)\\
\includegraphics[width=.5\textwidth,height=!]{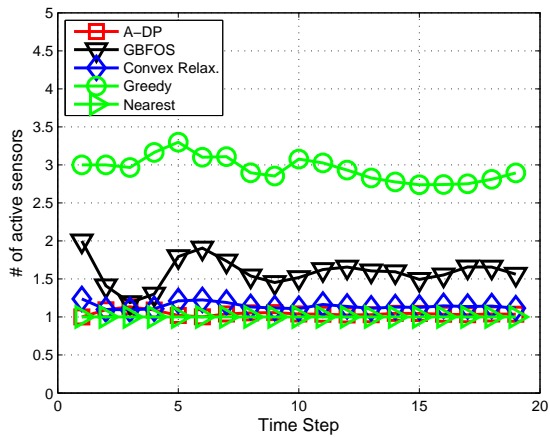} & \includegraphics[width=.5\textwidth,height=!]{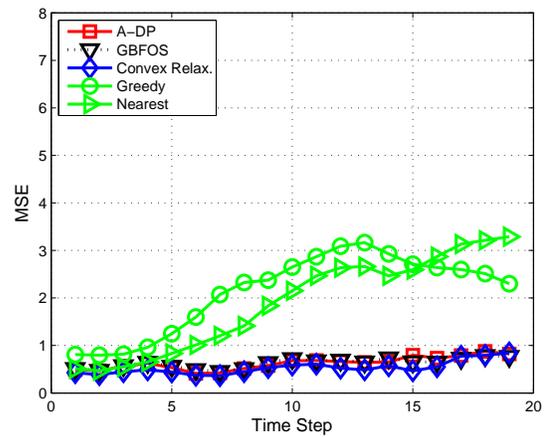}\\
(c) & (d)\\
\end{tabular}}
      \caption{ $N=25$, $R=5$, $T_{trial} = 500$  (a) Average number of active sensors,  $\rho = 2.5\times 10^{-3}$ , (b) MSE at each time step, $\rho = 2.5\times10^{-3}$,  (c) Average number of active sensors,  $\rho = 0.1$ (d) MSE at each time step, $\rho =0.1$.}
  \label{fig:Act_Sens_MSE_N25}
\end{figure*}

For $N=25$ sensors, Figs. \ref{fig:Act_Sens_MSE_N25}-(a) and \ref{fig:Act_Sens_MSE_N25}-(c) show the average number of sensors activated and Figs. \ref{fig:Act_Sens_MSE_N25}-(b) and \ref{fig:Act_Sens_MSE_N25}-(d) show the MSE at each time step of tracking. Since the sensor density is increased, the bandwidth allocation schemes tend to assign all the available bandwidth to a single sensor which has more precise information about the target. This improves the tracking performance at each time step. For $\rho = 2.5 \times 10^{-3}$ and $\rho = 0.1$ cases, convex relaxation, A-DP and GBFOS yield similar estimation performances and they significantly outperform the greedy search and nearest neighbor based bandwidth allocation approaches in terms of the MSE.

\section{Conclusion}
\label{sec5:Conclusions}

In this paper, we studied the dynamic bandwidth allocation problem for target tracking in a WSN with quantized
measurements. Under the bandwidth availability constraint,  we proposed two bandwidth distribution schemes which are based on convex relaxation and approximate DP to maximize the determinant of the FIM.  Simulation results show that convex relaxation, A-DP and GBFOS algorithms yield similar tracking performance, which is close to that provided by the optimal exhaustive search approach, and they outperform the greedy search and nearest neighbor approach significantly. Using the optimal solution of the convex optimization problem as the probability of transmission at each data rate, convex relaxation based bandwidth allocation satisfies the bandwidth constraint on the average while the other bandwidth distribution methods put a strict constraint on the bandwidth availability. In terms of computational complexity, A-DP is computationally more efficient than GBFOS and convex relaxation methods especially for a large sensor network with a large $N$.

In this work, we developed and compared bandwidth allocation schemes in target tracking for one step ahead only. Our future work will cover extensions of proposed schemes to non-myopic scenarios. Multi-target tracking by dynamic bandwidth allocation will also be considered as a future research direction.

\section*{Appendix}

It is easy to show that
\begin{equation}
\label{eq:app1}
\frac{\partial }{\partial x_t} Q\left(\frac{\eta_l^m - a_{i,t}}{\sigma}\right) = \frac{a_{i,t} n \alpha d_{i,t}^{n-2} (x_i - x_t)}{2\sqrt{2\pi \sigma^2} (1+\alpha d_{i,t}^n)} e^{-\frac{{(\eta^m_{l}-a_{i,t})}^2}{2\sigma^2}}
\end{equation} Then, substituting (\ref{eq:app1}) in (\ref{eq:app_lazim}), we have
\begin{eqnarray}
\label{eq:app2}
&& E \left[ - \frac{\partial^2 }{\partial x_t^2} p(D_{i,t} = l|\mathbf{x}_t, R_{i,t}=m) \right] \\
&& = \left( \sum_{l=0}^{2^m-1} \frac{\left[ e^{-\frac{{(\eta^m_{l}-a_{i,t})}^2}{2\sigma^2}}-e^{-\frac{{(\eta^m_{l+1}-a_{i,t})}^2}{2\sigma^2}}\right]^2}{ 8 \pi \sigma^2 p(D_{i,t} = l|\mathbf{x}_t, R_{i,t}=m)} \right) \frac{a_{i,t}^2 n^2 \alpha^2 d_{i,t}^{2n-4} (x_i - x_t)^2}{ (1+\alpha d_{i,t}^n)^2}  \nonumber \\
&& = \kappa_{i,t} (m,x_i,y_i,x_t,y_t) \frac{a_{i,t}^2 n^2 \alpha^2 d_{i,t}^{2n-4}}{(1+\alpha d_{i,t}^n)^2} (x_i - x_t)^2 \nonumber
\end{eqnarray}
Due to the symmetry between elements $x_t$ and $y_t$,
\begin{eqnarray}
\label{eq:app3}
& E \left[ - \frac{\partial^2 }{\partial y_t^2} p(D_{i,t} = l|\mathbf{x}_t, R_{i,t}=m) \right] & = \kappa_{i,t} (m,x_i,y_i,x_t,y_t) \frac{a_{i,t}^2 n^2 \alpha^2 d_{i,t}^{2n-4}}{(1+\alpha d_{i,t}^n)^2} (y_i - y_t)^2
\end{eqnarray}and
\begin{eqnarray}
\label{eq:app3}
&& E \left[ - \frac{\partial^2 }{\partial x_t \partial y_t} p(D_{i,t} = l|\mathbf{x}_t, R_{i,t}=m) \right]  \\
&& = \kappa_{i,t} (m,x_i,y_i,x_t,y_t) \frac{a_{i,t}^2 n^2 \alpha^2 d_{i,t}^{2n-4}}{(1+\alpha d_{i,t}^n)^2} (x_i - x_t)(y_i - y_t) \nonumber
\end{eqnarray}
%


\bibliography{dp_refs}
\bibliographystyle{IEEEtran}

\end{document}